\documentclass[a4paper,12pt]{article}

\usepackage[margin=1in,dvips,a4paper,nohead]{geometry}
\usepackage[margin=1em, font=small, justification=centerlast,
tableposition=top]{caption}

\usepackage{amsfonts,amssymb,amsmath}
\usepackage[noBBpl,slantedGreek]{mathpazo}

\usepackage{pstricks}

\psset{unit=1pt, dimen=middle, linewidth=.5pt, arrowsize=3pt 2}

\newdimen \psx
\newdimen \psy

\def\psddots (#1,#2){
  \psx #1\psxunit
  \psy #2\psyunit
  \qdisk(\psx,\psy){.7pt}
  \advance \psx -.2\psxunit
  \advance \psy .2\psyunit
  \qdisk(\psx,\psy){.7pt}
  \advance \psx .4\psxunit
  \advance \psy -.4\psyunit
  \qdisk(\psx,\psy){.7pt}
}

\def\pssddots (#1,#2){
  \psx #1\psxunit
  \psy #2\psyunit
  \qdisk(\psx,\psy){.7pt}
  \advance \psx .2\psxunit
  \advance \psy .2\psyunit
  \qdisk(\psx,\psy){.7pt}
  \advance \psx -.4\psxunit
  \advance \psy -.4\psyunit
  \qdisk(\psx,\psy){.7pt}
}

\makeatletter
\def\numberbysection{\@addtoreset{equation}{section}
        \def\theequation{\thesection.\arabic{equation}}}

\numberbysection

\def \:{\mskip .5\thinmuskip}
\def\ph {{\hbox to 0pt{\phantom{$\scriptstyle -1$}\hss}}}

\def \id{\mathop{\mathrm{id}}\nolimits}

\def\bbC {\mathbb C}

\def\bbR {\mathbb R}
\def\bbZ {\mathbb Z}
\def\calA {\mathcal A}

\def\calF {\mathcal F}
\def\calG {\mathcal G}
\def\calH {\mathcal H}
\def\calL {\mathcal L}
\def\calM {\mathcal M}
\def\calN {\mathcal N}
\def\calO {\mathcal O}
\def\calP {\mathcal P}

\mathchardef\Gamma "100
\def\gothg{\mathfrak g}
\def\gothG{\mathfrak G}
\def\gothgl{\mathfrak{gl}}
\def\gothH{\mathfrak H}

\def\d {\mathrm{d}}
\def\rme {\mathrm e}

\def\rmi {\mathrm i}
\def\rmGL {\mathrm{GL}}

\def\rmSO {\mathrm{SO}}
\def\wt {\widetilde}
\def\Mat {\mathrm{Mat}}

\title{\bf Abelian Toda solitons revisited}

\author{Kh. S. Nirov\\
\small \em Institute for Nuclear Research of the Russian Academy of
Sciences\\[-.3em]
\small \em 60th October Anniversary Prospect 7a, 117312 Moscow,
Russia\\[.3em]
A. V. Razumov\\
\small \em Institute for High Energy Physics\\[-.3em]
\small \em 142281 Protvino, Moscow Region, Russia}

\date{}

\begin{document}

\maketitle

\begin{abstract}
We present a systematic and detailed review of the application of 
the method of Hirota and the rational dressing method to abelian 
Toda systems associated with the untwisted loop groups of complex 
general linear groups. Emphasizing the rational dressing method, 
we compare the soliton solutions constructed within these two 
approaches, and show that the solutions obtained by the Hirota's 
method are a subset of those obtained by the rational dressing 
method.
\end{abstract}

\section{Introduction}

Two-dimensional Toda equations associated with loop
groups\footnote{Sometimes one deals with Toda equations associated
with affine groups being central extensions of loop groups. Usually it
is possible to construct solutions of the equations associated with
affine groups starting from solutions of the equations associated with
loop groups.} are very interesting examples of completely integrable
systems, see, for example, the monographs \cite{LezSav92,RazSav97}.
They possess soliton solutions having a nice physical interpretation
as interacting extended objects. Actually there is no a clear
definition of a soliton solution. In the present paper we call a
solution of equations an $n$-soliton solution, if it depends on $n$
linear combinations of independent variables.

Soliton solutions for Toda equations can be constructed with the help
of various methods. As far as we know, first explicit solutions of 
Toda equations associated with loop groups were found by Mikhailov 
\cite{Mik81}. He used the rational dressing method being a version 
of the inverse scattering method \cite{ZakSha79}. Note that in general 
the solutions obtained by Mikhailov are not soliton solutions. 
Besides, they are described by a redundant set of parameters.

Another method used here is the Hirota's one. Its essence \cite{Hir04}
is a change of the dependent variables which introduces the so called
$\tau$-functions. Here the final goal is to come to some special
bilinear partial differential equations which are solved then
perturbatively. The soliton solutions arise when the perturbation
series truncates at some finite order. This method was applied to
affine Toda systems, for example, in the papers \cite{Hol92,
ConFerGomZim93, MacMcG92, AraConFerGomZim93, ZhuCal93}. The main
disadvantage of the Hirota's method is that there is no a regular
method to find the desired transformation from the initial dependent
variables to $\tau$-functions. Therefore, sometimes it is used in
combination with other methods that helps to obtain a desired ansatz,
see, for example, the papers \cite{BueFerRaz02,AssFer07}.

There are also two additional approaches to the problem, being a
development of the Leznov--Saveliev method \cite{OliSavUnd93,
OliTurUnd92, OliTurUnd93}, and of the B\"{a}cklund--Darboux 
transformation 
\cite{ForGib80, MatSal91, LiaOliTur93, RogSch02, Zhou06}. 
These methods give the same soliton solutions as the Hirota's 
one and are not in the scope of the present paper.

Basic purpose of our review is to reproduce, in a possibly systematic
and detailed way, the application of the Hirota's and rational
dressing methods to Toda systems associated with the untwisted loop
groups of complex general linear groups, making an emphasis on the
rational dressing method, and compare the soliton solutions
constructed along these approaches. We show that all soliton solutions
obtained by the Hirota's method are contained among the solutions
obtained by the rational dressing method.

\section{Equations}

\subsection{Zero-curvature representation of Toda equations}

It is well known that Toda equations can be formulated as the zero-
curvature condition for a connection of a special form on the trivial
fiber bundle $\bbR^2 \times \calG \rightarrow \bbR^2$, where $\calG$
is a Lie group with the Lie algebra $\gothG$. The connection under
consideration can be identified with a $\gothG$-valued $1$-form
$\calO$ on $\bbR^2$. One can decompose such a connection over basis
$1$-forms as
\[
\calO = \calO_- {\mathrm{d}} z^- + \calO_+ {\mathrm{d}} z^+,
\]
where $z^-$, $z^+$ are the standard coordinates on the base manifold
$\bbR^2$, and the components $\calO_-$, $\calO_+$ are $\gothG$-valued
functions on it. Let us assume that the connection $\calO$ is flat
that means that its curvature is zero. This condition in terms of the
components has the form\footnote{We use the usual notation $\partial_-
= \partial/\partial z^-$ and $\partial_+ = \partial/\partial z^+$.}
\begin{equation}
\partial_- \calO_+ - \partial_+ \calO_- + [\calO_-, \calO_+] = 0.
\label{e:2.1}
\end{equation}
One can consider this relation as a system of partial differential
equations. In a sense, this system is trivial, and its general
solution is well known. It is given by the relations
\[
\calO_- = \Phi^{-1} \partial_- \Phi, \qquad
\calO_+ = \Phi^{-1} \partial_+ \Phi,
\]
where $\Phi$ is an arbitrary mapping of $\bbR^2$ to $\calG$. Actually
the triviality of the zero-cur\-va\-tu\-re condition is due to its
gauge invariance. That means that if a connection $\calO$ satisfies
(\ref{e:2.1}) then for an arbitrary mapping $\Psi$ of $\bbR^2$ to
$\calG$ the gauge transformed connection
\begin{equation}
\calO^\Psi = \Psi^{-1} \calO \Psi + \Psi^{-1} \d \Psi, \label{e:2.2}
\end{equation}
satisfies (\ref{e:2.1}) as well.

To obtain a nontrivial integrable system out of the zero-curvature
condition one imposes on the connection $\calO$ some restriction
destroying the gauge invariance. For the case of Toda equations they
are the grading and gauge fixing conditions which are introduced as
follows.

Suppose that the Lie algebra $\gothG$ is endowed with 
a $\bbZ$-gradation,
\[
\gothG = \bigoplus_{k \in \bbZ} \gothG_k,
\qquad
[\gothG_k , \gothG_l] \subset \gothG_{k+l},
\]
and that a positive integer $L$ is such that the grading subspaces
$\gothG_k$, where $0 < |k| < L$, are trivial.\footnote{It can be shown
that rejecting this restriction we do not come to new Toda equations,
see the paper~\cite{NirRaz07b}.} The grading condition states that the
components of $\calO$ have the form
\begin{equation}
\calO_- = \calO_{-0} + \calO_{-L}, \qquad \calO_+ = \calO_{+0} +
\calO_{+L}, \label{e:2.3}
\end{equation}
where $\calO_{-0}$ and $\calO_{+0}$ take values in $\gothG_0$, while
$\calO_{-L}$ and $\calO_{+L}$ take values in $\gothG_{-L}$ and
$\gothG_{+L}$ respectively. There is a residual gauge invariance.
Indeed, the gauge transformation (\ref{e:2.2}) with $\Psi$ taking
values in the Lie subgroup $\calG_0$ corresponding to the subalgebra
$\gothG_0$ does not violate the validity of the grading
condition~(\ref{e:2.3}). Therefore, one imposes an additional
condition, called the gauge fixing condition, of the form
\[
\calO_{+0} = 0.
\]
After that one can show that the components of the connection $\calO$
can be represented as
\begin{equation}
\calO_- = \Xi^{-1} \partial_- \Xi + \calF_-, \qquad \calO_+ = \Xi^{-1}
\calF_+ \Xi,
\label{e:2.4}
\end{equation}
where $\Xi$ is a mapping of $\bbR^2$ to $\calG_0$, $\calF_-$ and
$\calF_+$ are some mappings of $\bbR^2$ to $\gothG_{-L}$ and
$\gothG_{+L}$. One can easily get convinced that the zero-curvature
condition is equivalent to the equality\footnote{We assume for
simplicity that $\calG$ is a subgroup of the group formed by
invertible elements of some unital associative algebra $\calA$. In
this case $\gothG$ can be considered as a subalgebra of the Lie
algebra associated with $\calA$. Actually one can generalize our
consideration to the case of an arbitrary Lie group~$\calG$.}
\begin{equation}
\partial_+ (\Xi^{-1} \partial_- \Xi) = [\calF_-, \Xi^{-1} \calF_+ \Xi]
\label{e:2.5}
\end{equation}
and the relations
\begin{equation}
\partial_+ \calF_- = 0, \qquad \partial_- \calF_+ = 0.
\label{e:2.6}
\end{equation}
One supposes that the mappings $\calF_-$ and $\calF_+$ are fixed and
considers (\ref{e:2.5}) as an equation for $\Xi$ called the Toda
equation. When the group $\calG_0$ is abelian the corresponding Toda
equations are called abelian.

Thus, a Toda equation associated with a Lie group $\calG$ is specified
by a choice of a $\bbZ$-gradation of the Lie algebra $\gothG$ of
$\calG$ and mappings $\calF_-$, $\calF_+$ satisfying the
conditions~(\ref{e:2.6}). To classify the Toda equations associated
with a Lie group $\calG$ one should classify $\bbZ$-gradations of the
Lie algebra $\gothG$ of $\calG$.

Two remarks are in order. First, let $\Sigma$ be an isomorphism from a
$\bbZ$-graded Lie algebra $\gothG$ to a Lie algebra $\gothH$. One can
consider $\gothH$ as a $\bbZ$-graded Lie algebra with grading
subspaces $\gothH_k = \Sigma(\gothG_k)$. In such situation, one says
that $\mathbb Z$-gradations of $\gothG$ and $\gothH$ are conjugated by
$\Sigma$. Now let the Lie algebra $\gothG$ of the Lie group $\calG$ be
supplied with a $\bbZ$-gradation, and $\sigma$ be an isomorphism from
$\calG$ to a Lie group $\calH$. Denote by $\Sigma$ the isomorphism from 
the Lie algebra $\gothG$ to the Lie algebra $\gothH$ of the Lie group 
$\calH$ induced by the isomorphism $\sigma$. It is clear that if $\Xi$ 
is a solution of the Toda equation (\ref{e:2.5}), then the mapping
\begin{equation}
\Xi' = \sigma \circ \Xi \label{e:2.7}
\end{equation}
satisfies the Toda equation (\ref{e:2.5}) with the mappings $\calF_-$,
$\calF_+$ replaced by the mappings
\begin{equation}
\calF'_- = \Sigma \circ \calF_-, \qquad \calF'_+ = \Sigma \circ
\calF_+. \label{e:2.8}
\end{equation}
In other words, the solutions to two Toda equations under
consideration are connected via the isomorphism $\sigma$, and in this
sense these equations are equivalent. Thus, to describe really
different Toda equations it suffices to consider Lie groups and
$\bbZ$-gradations of their Lie algebras up to isomorphisms.

Secondly, let $\Theta_-$ and $\Theta_+$ be some mappings of $\bbR^2$
to $\calG_0$ which
satisfy the conditions
\[
\partial_+ \Theta_- = 0, \qquad \partial_- \Theta_+ = 0.
\]
If a mapping $\Xi$ satisfies the Toda equation (\ref{e:2.5}), then the
mapping
\begin{equation}
\Xi' = \Theta_+^{-1} \: \Xi \: \Theta_-, \label{e:2.9}
\end{equation}
satisfies the Toda equation (\ref{e:2.5}) where the mappings
$\calF_-$, $\calF_+$ are replaced by the mappings
\begin{equation}
\calF'_- = \Theta_-^{-1} \calF_- \Theta_-, \qquad \calF'_+ =
\Theta_+^{-1} \calF_+ \Theta^{}_+. \label{e:2.10}
\end{equation}
Again, the Toda equations for $\Xi$ and $\Xi'$ are not actually
different, and it is natural to use the transformations (\ref{e:2.10})
to make the mappings $\calF_-$, $\calF_+$ as simple as possible.

If the mappings $\Theta_-$ and $\Theta_+$ are such that
\[
\Theta_-^{-1} \calF_- \Theta_- = \calF_-, \qquad \Theta_+^{-1} \calF_+
\Theta^{}_+ = \calF_+.
\]
then the mapping $\Xi'$ satisfies the same Toda equation as the
mapping $\Xi$. Hence, in this case the transformation described by
relations (\ref{e:2.9}) is a symmetry transformation for the Toda
equation under consideration.

\subsection{Toda equations associated with loop groups of complex
simple Lie groups}

Let $\gothg$ be a finite dimensional real or complex Lie algebra. The
loop Lie algebra of $\gothg$, denoted $\calL(\gothg)$, is defined
alternatively either as the linear space $C^\infty(S^1, \gothg)$ of
smooth mappings of the circle $S^1$ to $\gothg$, or as the linear
space $C^\infty_{2 \pi}(\bbR, \gothg)$ of smooth $2 \pi$-periodic
mappings of the real line $\bbR$ to $\gothg$ with the Lie algebra
operation defined in both cases pointwise, see, for example,
\cite{PreSea86,Mil84,NirRaz06}. In this paper we adopt the second
definition and think of the circle $S^1$ as consisting of complex
numbers of modulus one. There is a convenient way to supply
$\calL(\gothg)$ with the structure of a Fr\'echet space, so that the
Lie algebra operation becomes a continuous mapping, see, for example,
\cite{Ham82,Mil84,NirRaz06}.

Now, let $G$ be a finite dimensional Lie group with the Lie algebra
$\gothg$. We define the loop group of $G$, denoted $\calL(G)$, as the
set $C^\infty(S^1, G)$ of smooth mappings of $S^1$ to $G$ with the
group law defined pointwise.  We assume that $\calL(G)$ is supplied
with the structure of a Fr\'echet manifold modeled on $\calL(\gothg)$
in such a way that it becomes a Lie group, see, for example,
\cite{Ham82,Mil84,NirRaz06}. The Lie algebra of the Lie group
$\calL(G)$ is naturally identified with the loop Lie
algebra~$\calL(\gothg)$.

Let $A$ be an automorphism of a finite dimensional Lie algebra
$\gothg$ satisfying the relation $A^M = \id_\gothg$ for some positive
integer $M$.\footnote{Note that we do not assume that $M$ is the order
of the automorphism $A$. It can be an arbitrary multiple of the
order.} The twisted loop Lie algebra $\calL_{A,M}(\gothg)$ is a
subalgebra of the loop Lie algebra $\calL(\gothg)$ formed by elements
$\xi$ that satisfy the equality
\[
\xi(\epsilon_M \bar{p}) = A(\xi(\bar{p})),
\]
where $\epsilon_M = \rme^{2 \pi \rmi/M}$ is the $M$th principal root
of unity. Similarly, given an automorphism $a$ of a Lie group $G$ that
satisfies the relation $a^M = \id_G$, we define the twisted loop group
$\calL_{a,M}(G)$ as the subgroup of the loop group $\calL(G)$ formed
by the elements $\chi$ satisfying the equality
\[
\chi(\epsilon_M \bar{p}) = a(\chi(\bar{p})).
\]
The Lie algebra of a twisted loop group $\calL_{a,M}(G)$ is naturally
identified with the twis\-ted loop Lie algebra $\calL_{A,M}(\gothg)$,
where we denote by $A$ the automorphism of the Lie algebra $\gothg$
corresponding to the automorphism $a$ of the Lie group $G$.

It is clear that loop groups and loop Lie algebras are partial cases
of twisted loop groups and twisted loop Lie algebras respectively.
Therefore, below by a loop group we mean either a usual loop group or
a twisted loop group, and by a loop Lie algebra we mean either a usual
loop Lie algebra or a twisted loop Lie algebra.

Now let us discuss the form of the Toda equations associated with a
loop group $\calL_{a,M}(G)$. First of all note that the group
$\calL_{a,M}(G)$ and its Lie algebra $\calL_{A,M}(\gothg)$ are
infinite dimensional manifolds. It appears that it is convenient to
reformulate the zero curvature representation of the Toda equations
associated with $\calL_{a,M}(G)$ in terms of finite dimensional
manifolds. To this end we use the so-called exponential law, see, for
example,~\cite{KriMic91, KriMic97}.

Let $\calM$, $\calN$, $\calP$ be three finite dimensional manifolds,
and $\calN$ be compact. Consider a smooth mapping $\calF$ of $\calM$
to $C^\infty(\calN, \calP)$. This mapping induces a mapping $f$ of
$\calM \times \calN$ to $\calP$ defined by the equality
\[
f(\bar{m}, \bar{n}) = (\calF(\bar{m}))(\bar{n}).
\]
It can be proved that the mapping $f$ is smooth. Conversely, if one
has a smooth mapping of $\calM \times \calN$ to $\calP$, reversing the
above equality one defines a mapping of $\calM$ to $C^\infty(\calN,
\calP)$, and this mapping is also smooth. Thus, we have the following
canonical identification
\[
C^\infty(\calM, C^\infty(\calN, \calP)) = C^\infty(\calM \times \calN,
\calP).
\]
It is this equality that is called the exponential law.

In the case under consideration the connection components $\calO_-$
and $\calO_+$ entering the equality (\ref{e:2.1}) are mappings of
$\bbR^2$ to the loop Lie algebra $\calL_{A,M}(\gothg)$. We will denote
the corresponding mappings of $\bbR^2 \times S^1$ to $\gothg$ by
$\omega_-$ and $\omega_+$, and call them also the connection
components. The mapping $\Phi$ generating the connection is a mapping
of $\bbR^2$ to $\calL_{a,M}(G)$. Denoting the corresponding mapping of
$\bbR^2 \times S^1$ by $\varphi$ we write
\begin{equation}
\varphi^{-1} \partial_- \varphi = \omega_-, \qquad \varphi^{-1}
\partial_+ \varphi = \omega_+. \label{e:2.11}
\end{equation}
Having in mind that the mapping $\varphi$ uniquely determines the
mapping $\Phi$, we say that the mapping $\varphi$ also generates the
connection under consideration.

The relations (\ref{e:2.4}) are equivalent to the equalities
\[
\omega_- = \gamma^{-1} \partial_- \gamma + f_-, \qquad \omega_+ =
\gamma^{-1} f_+ \gamma,
\]
where $\gamma$ is a smooth mapping of $\bbR^2 \times S^1$ to $G$
corresponding to the mapping $\Xi$, $f_-$ and $f_+$ are smooth
mappings of $\bbR^2 \times S^1$ to the Lie algebra $\gothg$ of $G$
corresponding to the mappings $\calF_-$ and $\calF_+$. The mappings
$f_-$ and $f_+$ satisfy the conditions
\begin{equation}
\partial_+ f_- = 0, \qquad \partial_- f_+ = 0, \label{e:2.12}
\end{equation}
which follow from the conditions (\ref{e:2.6}). The Toda equation
(\ref{e:2.5}) in the case under consideration is equivalent to the
equation
\begin{equation}
\partial_+(\gamma^{-1} \partial_- \gamma) = [f_-, \gamma^{-1} f_+
\gamma]. \label{e:2.13}
\end{equation}

To classify Toda equations associated with loop groups one has to
classify $\bbZ$-gra\-da\-ti\-ons of the corresponding loop Lie algebras.
This problem was partially solved in the paper \cite{NirRaz06}, see
also \cite{NirRaz07b}. In these papers the case of loop Lie algebras
of complex simple Lie algebras was considered and for this case a wide
class of the so-called integrable $\bbZ$-gradations \cite{NirRaz06}
with finite dimensional grading subspaces was described. Actually it
was shown that when $\gothg$ is a complex simple Lie algebra any
integrable $\mathbb Z$-gradation of a loop Lie algebra $\calL_{A,M}
(\gothg)$ with finite dimensional grading subspaces is conjugated by
an isomorphism to the standard gradation of another loop Lie algebra
$\calL_{A',M'}(\gothg)$, where the automorphisms $A$ and $A'$ differ
by an inner automorphism of $\gothg$. In particular, if $A$ is an
inner (outer) automorphism of $\gothg$, then $A'$ is also an inner
(outer) automorphism of~$\gothg$.

Assume now that $G$ is a finite dimensional complex simple Lie group,
then its Lie algebra $\gothg$ is a complex simple Lie algebra.
Consider a Toda equation associated with a loop group $\calL_{a, M}(G)
$. The corresponding $\bbZ$-gradation of $\calL_{A, M}(\gothg)$ is
conjugated by an isomorphism to the standard $\bbZ$-gradation of an
appropriate loop Lie algebra $\calL_{A',M'}(\gothg)$. Since the
automorphisms $A$ and $A'$ differ by an inner automorphism of
$\gothg$, the automorphism $A'$ can be lifted to an automorphism $a'$
of $G$, and the isomorphism from $\calL_{A, M}(\gothg)$ to $\calL_{A',
M'}(\gothg)$ under consideration can be lifted to an isomorphism from
$\calL_{a, M}(G)$ to $\calL_{a', M'}(G)$. Actually this means that the
initial Toda equation associated with $\calL_{a, M}(G)$ is equivalent
to a Toda equation associated with $\calL_{a', M'}(G)$ arising when we
supply $\calL_{A', M'}(\gothg)$ with the standard $\bbZ$-gradation.

The grading subspaces for the standard $\bbZ$-gradation of a loop Lie
algebra $\calL_{A, M}(\gothg)$ are
\[
\calL_{A, M}(\gothg)_{k} = \{ \xi \in \calL_{A,M}(\gothg) \mid \xi =
\lambda^k x, \ A(x) = \epsilon_M^{k} x \},
\]
where by $\lambda$ we denote the restriction of the standard
coordinate on $\bbC$ to $S^1$. It is very useful to realize that every
automorphism $A$ of the Lie algebra $\gothg$ satisfying the relation
$A^M = \mathrm{id}_\gothg$ induces a $\mathbb Z_M$-gradation of
$\gothg$ with the grading subspaces\footnote{We denote by $[k]_M$ the
element of the ring $\mathbb Z_M$ corresponding to the integer $k$.}
\[
\gothg_{[k]_M} = \{x \in \gothg \mid A(x) = \epsilon_M^{k} x\}, \qquad
k = 0, \ldots, M-1.
\]
Vice versa, any $\mathbb Z_M$-gradation of $\gothg$ defines in an
evident way an automorphism $A$ of $\gothg$ satisfying the relation
$A^M = \mathrm{id}_\gothg$. A $\mathbb Z_M$-gradation of $\gothg$ is
called an inner or outer type gradation, if the associated
automorphism $A$ of $\gothg$ is of inner or outer type respectively.
In terms of the corresponding $\bbZ_M$-gradation the grading subspaces
for the standard $\bbZ$-gradation of a loop Lie algebra $\calL_{A, M}
(\gothg)$ are
\[
\calL_{A, M}(\gothg)_{k} = \{ \xi \in \calL_{A,M}(\gothg) \mid \xi =
\lambda^k x, \ x \in \gothg_{[k]_M} \}.
\]

It is evident that for the standard $\bbZ$-gradation the subalgebra
$\calL_{A, M}(\gothg)_0$ is isomorphic to the subalgebra 
$\gothg_{[0]_M}$ of $\gothg$, and the Lie group $\calL_{a, M}(G)_0$ 
is isomorphic to the connected Lie subgroup $G_0$ of $G$ corresponding 
to the Lie algebra $\gothg_{[0]_M}$. Hence, the mapping $\gamma$ is 
actually a mapping of $\bbR^2$ to $G_0$. The mappings $f_-$ and $f_+$ 
are given by the relation
\[
f_-(\bar{m}, \bar{p}) = \bar{p}^{-L} c_-(\bar{m}), \qquad 
f_+(\bar{m}, \bar{p}) = \bar{p}^L c_+(\bar{m}), \qquad 
\bar{m} \in \bbR^2, \quad \bar{p} \in S^1,
\]
where $c_-$ and $c_+$ are mappings of $\bbR^2$ to $\gothg_{-[L]_M}$
and $\gothg_{+[L]_M}$ respectively. For the connection components
$\omega_-$ and $\omega_+$ we have
\begin{equation}
\omega_- = \gamma^{-1} \partial_- \gamma + \lambda^{-L} c_-, \qquad
\omega_+ = \lambda^L \gamma^{-1} c_+ \gamma, \label{e:2.14}
\end{equation}
and the Toda equation (\ref{e:2.13}) can be written as
\begin{equation}
\partial_+(\gamma^{-1} \partial_- \gamma) = [c_-, \gamma^{-1} c_+
\gamma]. \label{e:2.15}
\end{equation}
The conditions (\ref{e:2.12}) imply that
\begin{equation}
\partial_+ c_- = 0, \qquad \partial_- c_+ = 0. \label{e:2.16}
\end{equation}
It is natural to call an equation of the form (\ref{e:2.15}) also a
Toda equation.

Let $b$ be an automorphism of the Lie group $G$ and $B$ be the
corresponding automorphism of the Lie algebra $\gothg$. The mapping
$\sigma$ defined by the equality
\[
\sigma (\chi) = b \circ \chi, \qquad \chi \in \calL_{a, M}(G),
\]
is an isomorphism from $\calL_{a, M}(G)$ to $\calL_{a', M}(G)$, where
$a'$ is an automorphism of $G$ defined as
\[
a' = b \: a \: b^{-1}.
\]
It is clear that the mapping $\Sigma$ defined by the equality
\[
\Sigma (\xi) = B \circ \xi, \qquad \xi \in \calL_{A, M}(\gothg),
\]
is an isomorphism from $\calL_{A, M}(\gothg)$ to $\calL_{A', M}
(\gothg)$, where $A'$ is an automorphism of $\gothg$ corresponding to
the automorphism $a'$ of $G$. With such isomorphism in the case under
consideration the transformations (\ref{e:2.7}) and (\ref{e:2.8}) take
the form
\begin{gather*}
\gamma' = b \circ \gamma, \\
c'_- = B \circ c_-, \qquad c'_+ = B \circ c_+.
\end{gather*}
If a mapping $\gamma$ satisfies the Toda equations (\ref{e:2.15}),
then the mapping $\gamma'$ satisfies the Toda equation (\ref{e:2.15})
where the mappings $c_-$, $c_+$ are replaced by the mappings $c'_-$
and~$c'_+$. Actually this means that Toda equations of the form
(\ref{e:2.15}) defined by means of conjugated $\bbZ_M$-gradations are
equivalent.

The transformations (\ref{e:2.9}) and (\ref{e:2.10}) take now the
forms
\begin{gather}
\gamma' = \eta_+^{-1} \: \gamma \: \eta_-, \label{e:2.17} \\
c'_- = \eta_-^{-1} c_- \eta_-, \qquad c'_+ = \eta_+^{-1} c_+ \eta^{}
_+, \label{e:2.18}
\end{gather}
where $\eta_-$ and $\eta_+$ are some mappings of $\bbR^2 \times S^1$
to $G_0$ that satisfy the conditions
\begin{equation}
\partial_+ \eta_- = 0, \qquad \partial_- \eta_+ = 0. \label{e:2.19}
\end{equation}
Again, if a mapping $\gamma$ satisfies the Toda equations
(\ref{e:2.15}), then the mapping $\gamma'$ satisfies the Toda equation
(\ref{e:2.15}) where the mappings $c_-$, $c_+$ are replaced by the
mappings $c'_-$ and~$c'_+$. If the mappings $\eta_-$ and $\eta_+$ are
such that
\begin{equation}
\eta_-^{-1} c_- \eta_- = c_-, \qquad \eta_+^{-1} c_+ \eta^{}_+ = c_+
\label{e:2.20}
\end{equation}
then the transformation (\ref{e:2.17}) is a symmetry transformation
for the Toda equation under consideration.

Thus, if $G$ is a finite dimensional complex simple Lie group, then
the Toda equation associated with a loop group $\calL_{a, M}(G)$ and
defined with the help of an integrable $\bbZ$-gradation of $\calL_{A,
M}(\gothg)$ with finite dimensional grading subspaces is equivalent to
an equation of the form (\ref{e:2.15}). To describe all nonequivalent
Toda equations of this type one has to classify finite order
automorphisms of the Lie algebra $\gothg$ or, equivalently, its
$\bbZ_M$-gradations up to conjugations.\footnote{Strictly speaking, we
have to consider only conjugations by automorphisms of $\gothg$ which
can be lifted to automorphisms of $G$.} This problem was solved quite
a long time ago, see, for example, \cite{Kac94, GorOniVin94}. However,
it appeared that the classification described in \cite{Kac94,
GorOniVin94} is not convenient for classification of Toda equations.
Restricting to the case of loop Lie algebras of complex classical Lie
algebras one can use another classification based on a convenient
block matrix representation of the grading subspaces \cite{NirRaz07a,
NirRaz07b}. Let us describe the main points of the resulting
classification of Toda equations.

Each element $x$ of the complex classical Lie algebra $\gothg$ under
consideration is considered as a $p \times p$ block matrix $(x_{\alpha
\beta})$, where $x_{\alpha \beta}$ is an $n_\alpha \times n_\beta$
matrix. Certainly, the sum of the positive integers $n_\alpha$ is the
size $n$ of the matrices representing the elements of $\gothg$. For
the case of Toda systems associated with the loop groups $\calL_{a, M}
(\rmGL_n(\bbC))$, where $a$ is an inner automorphism of
$\rmGL_n(\bbC)$, the integers $n_\alpha$ are arbitrary. For the other
cases they should satisfy some restrictions dictated by the structure
of the Lie algebra $\gothg$.

The mapping $\gamma$ has the block diagonal form
\[
\psset{xunit=1.7em, yunit=1.2em}
\gamma = \left( \raise -1.8\psyunit \hbox{\begin{pspicture}(.6,.6)
(4.5,4.2)
\rput(1,4){$\Gamma_1$}
\rput(2,3){$\Gamma_2$}
\qdisk(2.7,2.3){.7pt} \qdisk(3,2){.7pt} \qdisk(3.3,1.7){.7pt}
\rput(4,1){$\Gamma_p$}
\end{pspicture}} \right).
\]
For each $\alpha = 1, \ldots, p$ the mapping $\Gamma_\alpha$ is a
mapping of $\bbR^2$ to the Lie group $\rmGL_{n_\alpha}(\bbC)$. For the
case of Toda systems associated with the loop groups $\calL_{a, M}
(\rmGL_n(\bbC))$, where $a$ is an inner automorphism of
$\rmGL_n(\bbC)$, the mappings $\Gamma_\alpha$ are arbitrary. For the
other cases they satisfy some additional restrictions.

The mapping $c_+$ has the following block matrix structure:
\[
\psset{xunit=2.5em, yunit=1.4em}
c_+ = \left( \raise -2.4\psyunit \hbox{\begin{pspicture}(.6,.5)
(5.6,5.3)
\rput(1,5){$0$} \rput(2,4.92){$C_{+1}$}
\rput(2,4){$0$}
\qdisk(2.8,4.2){.7pt} \qdisk(3,4){.7pt} \qdisk(3.2,3.8){.7pt}
\qdisk(2.8,3.2){.7pt} \qdisk(3,3){.7pt} \qdisk(3.2,2.8){.7pt}
\qdisk(3.8,3.2){.7pt} \qdisk(4,3){.7pt} \qdisk(4.2,2.8){.7pt}
\rput(4,2){$0$} \rput(5,1.87){$C_{+(p-1)}$}
\rput(1,.94){$C_{+0}$} \rput(5,1){$0$}
\end{pspicture}} \right),
\]
where for each $\alpha = 1, \ldots, p-1$ the mapping $C_{+\alpha}$ is
a mapping of $\bbR^2$ to the space of $n_\alpha \times n_{\alpha+1}$
complex matrices, and $C_{+0}$ is a mapping of $\bbR^2$ to the space
of $n_p \times n_1$ complex matrices. The mapping $c_-$ has a similar
block matrix structure:
\[
\psset{xunit=2.5em, yunit=1.4em}
c_- = \left( \raise -2.4\psyunit \hbox{\begin{pspicture}(.5,.5)
(5.5,5.3)
\rput(1,5){$0$} \rput(5,4.92){$C_{-0}$}
\rput(1,4){$C_{-1}$} \rput(2,4){$0$}
\qdisk(1.8,3.2){.7pt} \qdisk(2,3){.7pt} \qdisk(2.2,2.8){.7pt}
\qdisk(2.8,3.2){.7pt} \qdisk(3,3){.7pt} \qdisk(3.2,2.8){.7pt}
\qdisk(2.8,2.2){.7pt} \qdisk(3,2){.7pt} \qdisk(3.2,1.8){.7pt}
\rput(4,1.87){$0$}
\rput(4,.94){$C_{-(p-1)}$} \rput(5,1){$0$}
\end{pspicture}} \right),
\]
where for each $\alpha = 1, \ldots, p-1$ the mapping $C_{-\alpha}$ is
a mapping of $\calM$ to the space of $n_{\alpha+1} \times n_\alpha$
complex matrices, and $C_{-0}$ is a mapping of $\calM$ to the space of
$n_1 \times n_p$ complex matrices. The conditions (\ref{e:2.16}) imply
\[
\partial_+ C_{-\alpha} = 0, \qquad \partial_- C_{+\alpha} = 0, \qquad
\alpha = 0, 1, \ldots, p-1.
\]
For the case of Toda systems associated with the loop groups
$\calL_{a, M}(\rmGL_n(\bbC))$, where $a$ is an inner automorphism of
$\rmGL_n(\bbC)$, the mappings $C_{\pm \alpha}$ are arbitrary. For the
other cases they should satisfy some additional restrictions.

It is not difficult to show that the Toda equation (\ref{e:2.15}) is
equivalent to the following system of equations for the mappings
$\Gamma_\alpha$:
\begin{align}
\partial_+ \left( \Gamma_1^{-1} \: \partial_- \Gamma_1^{} \right)
&= - \Gamma_1^{-1} C_{+1}^{} \: \Gamma_2^{} \: C_{-1}^{}
+ C_{-0}^{} \Gamma_p^{-1} C_{+0}^{} \Gamma_1^{},
\notag \\*
\partial_+ \left( \Gamma_2^{-1} \: \partial_- \Gamma_2^{} \right)
&= - \Gamma_2^{-1} C_{+2}^{} \: \Gamma_3^{} \: C_{-2}^{}
+ C_{-1}^{} \Gamma_1^{-1} C_{+1}^{} \Gamma_2^{},
\notag \\*
& \quad \vdots
\label{e:2.21} \\*
\partial_+ \left(\Gamma_{p-1}^{-1} \: \partial_-
\Gamma_{p-1}^{}\right)
&= - \Gamma_{p-1}^{-1} C_{+(p-1)}^{} \: \Gamma_p^{} \: C_{-(p-1)}^{}
+ C_{-(p-2)}^{} \Gamma_{p-2}^{-1} C_{+(p-2)}^{} \Gamma_{p-1}^{},
\notag \\*
\partial_+ \left( \Gamma_p^{-1} \: \partial_- \Gamma_p^{} \right)
&= - \Gamma_p^{-1} C_{+0}^{} \: \Gamma_1^{} \: C_{-0}^{}
+ C_{-(p-1)}^{} \Gamma_{p-1}^{-1} C_{+(p-1)}^{} \Gamma_p^{}.
\notag
\end{align}
It appears that in the case under consideration without any loss of
generality one can assume that the positive integer $L$, entering the
construction of Toda equations, is equal to $1$. Note also that if any
of the mappings $C_{+\alpha}$ or $C_{-\alpha}$ is a zero mapping, then
the equations (\ref{e:2.21}) are actually equivalent to a Toda
equation associated with a finite dimensional group or to a set of two
such equations.

\subsection{Abelian Toda equations associated with loop groups of
complex general linear Lie groups}

There are three types of abelian Toda equations associated with
$\calL_{a, M}(\mathrm{GL}_n(\bbC))$.

\subsubsection{First type} \label{s:2.3.1}

The abelian Toda equations of the first type arise when the
automorphism $A$ is defined by the equality
\[
A(x) = h x h^{-1}, \qquad x \in \gothgl_n(\bbC),
\]
where $h$ is a diagonal matrix with the diagonal matrix elements
\begin{equation}
h_{kk} = \epsilon_n^{n-k+1}, \qquad k = 1, \ldots, n. \label{e:2.22}
\end{equation}
The corresponding automorphism $a$ of $\rmGL_n(\bbC)$ is defined by
the equality
\begin{equation}
a(g) = h g h^{-1}, \qquad g \in \rmGL_n(\bbC), \label{e:aut}
\end{equation}
where again $h$ is a diagonal matrix determined by the relation
(\ref{e:2.22}). Here the integer $M$ is equal to $n$, and $A$ is an
inner automorphism which generates a $\bbZ_n$-gradation of
$\gothgl_n(\bbC)$. The block matrix structure related to this
gradation is the matrix structure itself. In other words, all blocks
are of size one by one. The mappings $\Gamma_\alpha$ are mappings of
$\bbR^2$ to the Lie group $\rmGL_1(\bbC)$ which is isomorphic to the
Lie group $\bbC^\times = \bbC \smallsetminus \{0\}$. The mappings
$C_{\pm \alpha}$ are just complex functions on $\bbR^2$. The Toda
equations under consideration have the form (\ref{e:2.21}) with $p =
n$.

Let us describe the action of the transformations (\ref{e:2.17}) and
(\ref{e:2.18}) on the equations (\ref{e:2.21}) in the case under
consideration. The mappings $\eta_-$ and $\eta_+$ have a diagonal form
and we denote
\[
(\eta_-)_{\alpha \alpha} = H_{-\alpha}, \qquad (\eta_+)_{\alpha
\alpha} = H_{+\alpha}, \qquad \alpha = 1, \ldots, n.
\]
The functions $H_{-\alpha}$ and $H_{+\alpha}$ satisfy the relations
\[
\partial_+ H_{-\alpha} = 0, \qquad \partial_- H_{+\alpha} = 0,
\]
which follow from the relations (\ref{e:2.19}). In terms of the
functions $C_{-\alpha}$ and $C_{+\alpha}$ the transformations
(\ref{e:2.17}) and (\ref{e:2.18}) look as
\begin{gather}
\Gamma'_\alpha = H_{+\alpha}^{-1} \Gamma_\alpha^\ph H_{-\alpha}^\ph,
\label{e:2.23} \\
C'_{-\alpha} = H_{-(\alpha+1)}^{-1} C_{-\alpha}^\ph H_{-\alpha}^\ph,
\qquad
C'_{+\alpha} = H_{+\alpha}^{-1} C_{+\alpha}^\ph H_{+(\alpha+1)}^\ph.
\label{e:2.24}
\end{gather}
Assume that the functions $C_{-\alpha}$ and $C_{+\alpha}$ have no
zeros. Let us show that in this case the functions $H_{-\alpha}$ and
$H_{+\alpha}$ can be chosen in such a way that $C'_{-\alpha} =
C_-^\ph$ and $C'_{+\alpha} = C_+^\ph$ for some functions $C_-$ and
$C_+$ which have no zeros and are subject to the conditions
\begin{equation}
\partial_+ C_- = 0, \qquad \partial_- C_+ = 0. \label{e:2.25}
\end{equation}
Indeed, let $C_-$ and $C_+$ be some functions which satisfy the
equalities
\[
C_-^n = \prod_{\alpha = 1}^n C_{-\alpha}, \qquad C_+^n = \prod_{\alpha
= 1}^n C_{+\alpha}.
\]
One can verify that the transformations (\ref{e:2.24}) with
\[
H_{-\alpha} = \prod_{\beta = \alpha}^n \frac{C_-}{C_{-\beta}},
\qquad H_{+\alpha} = \prod_{\beta = \alpha}^n \frac{C_{+\beta}}{C_+}
\]
give the desired result, $C'_{-\alpha} = C_-^\ph$ and $C'_{+\alpha} =
C_+^\ph$. The methods to find soliton solutions described below work
for arbitrary functions $C_-$ and $C_+$. However, to simplify formulas
we will only consider the case when $C_- = m$, and $C_+ = m$, where
$m$ is a nonzero constant. In other words, we will assume that
\begin{equation}
\psset{xunit=1.6em, yunit=1.2em}
c_- = m \left( \raise -2.4\psyunit \hbox{\begin{pspicture}(.5,.5)
(5.5,5.3)
\rput(1,5){$0$} \rput(5,5){$1$}
\rput(1,4){$1$} \rput(2,4){$0$}
\qdisk(1.7,3.3){.7pt} \qdisk(2,3){.7pt} \qdisk(2.3,2.7){.7pt}
\qdisk(2.7,3.3){.7pt} \qdisk(3,3){.7pt} \qdisk(3.3,2.7){.7pt}
\qdisk(2.7,2.3){.7pt} \qdisk(3,2){.7pt} \qdisk(3.3,1.7){.7pt}
\rput(4,2){$0$}
\rput(4,1){$1$} \rput(5,1){$0$}
\end{pspicture}} \right), \qquad
c_+ = m \left( \raise -2.4\psyunit \hbox{\begin{pspicture}(.6,.5)
(5.6,5.3)
\rput(1,5){$0$} \rput(2,5){$1$}
\rput(2,4){$0$}
\qdisk(2.7,4.3){.7pt} \qdisk(3,4){.7pt} \qdisk(3.3,3.7){.7pt}
\qdisk(2.7,3.3){.7pt} \qdisk(3,3){.7pt} \qdisk(3.3,2.7){.7pt}
\qdisk(3.7,3.3){.7pt} \qdisk(4,3){.7pt} \qdisk(4.3,2.7){.7pt}
\rput(4,2){$0$} \rput(5,2){$1$}
\rput(1,.94){$1$} \rput(5,1){$0$}
\end{pspicture}} \right). \label{e:2.26}
\end{equation}
The equations under consideration take in this case the form
\begin{align}
\partial_+ \left( \Gamma_1^{-1} \: \partial_- \Gamma_1^\ph \right)
&= - m^2 (\Gamma_1^{-1} \Gamma_2^\ph - \Gamma_p^{-1} \Gamma_1^\ph),
\notag \\*
\partial_+ \left( \Gamma_2^{-1} \: \partial_- \Gamma_2^{} \right)
&= - m^2 (\Gamma_2^{-1} \Gamma_3^\ph - \Gamma_1^{-1} \Gamma_2^\ph),
\notag \\*
& \quad \vdots \label{e:2.27} \\*
\partial_+ \left(\Gamma_{n-1}^{-1} \: \partial_- \Gamma_{n-1}^{}
\right)
&= - m^2 (\Gamma_{n-1}^{-1} \Gamma_n^\ph - \Gamma_{n-2}^{-1}
\Gamma_{n-1}^\ph), \notag \\*
\partial_+ \left( \Gamma_n^{-1} \: \partial_- \Gamma_n^{} \right)
&= - m^2 (\Gamma_n^{-1} \Gamma_1^\ph - \Gamma_{n-1}^{-1}
\Gamma_n^\ph). \notag
\end{align}
It is worth to note that when the functions $C_-$ and $C_+$ are real
one can come to the Toda equations with $C_- = m$ and $C_+ = m$ by an
appropriate change of the coordinates $z^-$ and $z^+$.

The symmetry transformations (\ref{e:2.20}) for the system under
consideration are described by the relations (\ref{e:2.23}) where
$H_{-\alpha} = H_-$ and $H_{+\alpha} = H_+$ for some
functions $H_-$ and $H_+$ satisfying the conditions
\[
\partial_+ H_- = 0, \qquad \partial_- H_+ = 0.
\]
In particular, multiplication of all $\Gamma_\alpha$ by the same
constant is a symmetry transformation.

Defining
\[
\Gamma = \Gamma_1 \Gamma_2 \ldots \Gamma_n,
\]
one can easily see that
\[
\partial_+(\Gamma^{-1} \partial_- \Gamma) = \sum_{\alpha=1}^n
\partial_+(\Gamma_\alpha^{-1} \partial_- \Gamma_\alpha^\ph).
\]
Equations (\ref{e:2.27}) give
\[
\partial_+(\Gamma^{-1} \partial_- \Gamma) = 0,
\]
therefore,
\[
\Gamma = \Gamma_+ \Gamma_-^{-1},
\]
for some functions $\Gamma_-$ and $\Gamma_+$ which satisfy the
relations
\[
\partial_+ \Gamma_- = 0, \qquad \partial_- \Gamma_+ = 0.
\]
Thus, if we perform the symmetry transformation (\ref{e:2.23}) with
$H_{-\alpha}$ and $H_{+\alpha}$ given by
\[
H_{-\alpha} = \Gamma_-^{1/n}, \qquad H_{+\alpha} = \Gamma_+^{1/n},
\]
we will obtain functions $\Gamma'_i$ which satisfy the Toda equations
(\ref{e:2.27}) and obey the equality
\[
\Gamma' = \Gamma'_1 \Gamma'_2 \ldots \Gamma'_n = 1.
\]
Actually this means that via appropriate symmetry transformations we
can reduce solutions of the abelian Toda equations associated with the
loop group of $\mathrm{GL}_n(\bbC)$ under consideration to solutions
of the corresponding Toda equations associated with the loop group of
$\mathrm{SL}_n(\bbC)$.

Suppose now that we have a solution of the equations (\ref{e:2.27})
with $\Gamma = 1$. The mappings $\Gamma_\alpha$ for $\alpha = 1,
\ldots, n-1$ are independent. Introduce a new set of $n-1$ independent
mappings $\Phi_\alpha$, $\alpha = 1, \ldots, n - 1$, defined as
\[
\Phi_\alpha = \prod_{\beta=1}^\alpha \Gamma_\beta.
\]
It is easy to show that the inverse transition to the mappings
$\Gamma_\alpha$ is described by the equalities
\[
\Gamma_1 = \Phi_1, \quad \Gamma_2 = \Phi_1^{-1} \Phi_2, \quad \ldots
\quad
\Gamma_{n-1} = \Phi_{n-2}^{-1} \Phi_{n-1}, \quad \Gamma_n = \Phi_{n-1}
^{-1},
\]
and that the mappings $\Phi_\alpha$ satisfy the equations
\begin{align*}
\partial_+(\Phi_1^{-1} \partial_- \Phi_1) &= - m^2 (\Phi_1^{-2} \Phi_2
-
\Phi_{n-1} \Phi_1), \\
\partial_+(\Phi_2^{-1} \partial_- \Phi_2) &= - m^2 (\Phi_1 \Phi_2^{-2}
\Phi_3 - \Phi_{n-1} \Phi_1), \\
& \vdots \\
\partial_+(\Phi_{n-2}^{-1} \partial_- \Phi_{n-2}) &= - m^2 (\Phi_{n-3}
\Phi_{n-2}^{-2} \Phi_{n-1} - \Phi_{n-1} \Phi_1), \\
\partial_+(\Phi_{n-1}^{-1} \partial_- \Phi_{n-1}) &= - m^2 (\Phi_{n-2}
\Phi_{n-1}^{-2} - \Phi_{n-1} \Phi_1).
\end{align*}
This system can be written in a more symmetric form. To this end one
introduces an additional mapping $\Delta_0$, which satisfies the
equation
\[
\partial_+(\Delta_0^{-1} \partial_- \Delta_0) = - m^2 \Phi_{n-1}
\Phi_1,
\]
and denotes
\[
\Delta_\alpha = \Delta_0 \Phi_\alpha, \qquad \alpha = 1, \ldots, n-1.
\]
It is easy to see that the mappings $\Delta_\alpha$, $\alpha = 0, 1,
\ldots, n-1$, satisfy the equations
\begin{align}
\partial_+(\Delta_0^{-1} \partial_- \Delta_0) &= - m^2 \Delta_{n-1}
\Delta_0^{-2} \Delta_1, \notag \\
\partial_+(\Delta_1^{-1} \partial_- \Delta_1) &= - m^2 \Delta_0
\Delta_1^{-2} \Delta_2, \notag \\
& \vdots \label{e:2.28} \\
\partial_+(\Delta_{n-2}^{-1} \partial_- \Delta_{n-2}) &= - m^2
\Delta_{n-3} \Delta_{n-2}^{-2} \Delta_{n-1}, \notag \\
\partial_+(\Delta_{n-1}^{-1} \partial_- \Delta_{n-1}) &= - m^2
\Delta_{n-2} \Delta_{n-1}^{-2} \Delta_0, \notag
\end{align}
which can be written as
\begin{equation}
\partial_+ (\Delta_\alpha^{-1} \partial_- \Delta_\alpha) = - m^2
\prod_{\beta=0}^{n-1}
\Delta^{-a_{\alpha \beta}}_\beta, \label{e:2.29}
\end{equation}
where $a_{\alpha \beta}$ are the matrix elements of the Cartan matrix
of an affine
Lie algebra of type $A^{(1)}_{n-1}$:
\begin{equation}
(a_{\alpha \beta}) = \left( \begin{array}{rrrcrrr}
2 & -1 & 0 & \cdots & 0 & 0 & -1 \\
-1 & 2 & -1 & \cdots & 0 & 0 & 0 \\
0 & -1 & 2 & \cdots & 0 & 0 & 0 \\
\vdots & \vdots & \vdots & \ddots & \vdots & \vdots & \vdots \\
0 & 0 & 0 & \cdots & 2 & -1 & 0 \\
0 & 0 & 0 & \cdots & -1 & 2 & -1 \\
-1 & 0 & 0 & \cdots & 0 & -1 & 2
\end{array} \right). \label{e:2.30}
\end{equation}
The equations (\ref{e:2.29}) are of the standard form of the Toda
equations associated with the $A_{n-1}^{(1)}$ affine Lie group.

\subsubsection{Second type}

The abelian Toda equations of the second and third types arise when we
use outer automorphisms of $\gothgl_n(\bbC)$. For the equations of the
second type $n$ is odd, and for the equations of the third type $n$ is
even.

Consider first the case of an odd $n = 2s - 1$, $s \ge 2$. In this
case an abelian Toda equation arises when the automorphism $A$ is
defined by the equality
\begin{equation}
A(x) = - h (B^{-1} {}^{t\!} x B) h^{-1}, \label{e:2.31}
\end{equation}
where ${}^{t\!} x$ means the transpose of $x$, $h$ is a diagonal
matrix with the diagonal matrix elements
\[
h_{kk} = \epsilon_{2n}^{n-k+1} = \epsilon_{4s-2}^{2s-k},
\]
and $B$ is an $n \times n$ matrix of the form
\[
\psset{xunit=1.6em, yunit=1em}
B = \left( \raise -3.3\psyunit \hbox{\begin{pspicture}(.7,.5)(7.3,7.1)
\rput(1,7){$1$} \rput(7,6){$1$}
\qdisk(6.3,5.3){.7pt} \qdisk(6,5){.7pt} \qdisk(5.7,4.7){.7pt}
\rput(5,4){$1$} \rput(4,3){$-1$}
\qdisk(3.3,2.3){.7pt} \qdisk(3,2){.7pt} \qdisk(2.7,1.7){.7pt}
\rput(2,1){$-1$}
\psline(.7,6.5)(7.3,6.5)
\psline(1.5,7.3)(1.5,.5)
\end{pspicture}} \right).
\]
The order of the automorphism $A$ is $2n = 4s-2$ and the integer $p$
is $2s - 1$. The mapping $\gamma$ is a diagonal matrix, and the
mappings $\Gamma_\alpha$ are mappings of $\bbR^2$ to $\bbC^\times$
subject to the constraints
\[
\Gamma_1 = 1, \qquad \Gamma_{2s-\alpha+1} = \Gamma_\alpha^{-1}, \quad
\alpha = 2, \ldots, s.
\]
The mappings $C_{\pm \alpha}$ are complex functions satisfying the
equality
\begin{equation}
C_{\pm 0} = C_{\pm 1}, \label{e:2.32}
\end{equation}
and for $s > 2$ the equalities
\begin{equation}
C_{\pm (2s - \alpha)} = - C_{\pm \alpha}, \quad \alpha = 2, \ldots,
s-1. \label{e:2.33}
\end{equation}
Let us choose the mappings $\Gamma_\alpha$, $\alpha = 2, \ldots, s$,
as a complete set of mappings parameterizing the mapping $\gamma$.
Taking into account the equalities (\ref{e:2.32}) and (\ref{e:2.33})
we come to the following set of independent equations equivalent to
the Toda equation under consideration
\begin{align}
\partial_+(\Gamma_2^{-1} \partial_- \Gamma_2^{\ph}) & = - C_{+2}^{\ph}
C_{-2}^{\ph} \Gamma_2^{-1} \Gamma_3^{\ph} + C_{+1}^{\ph} C_{-1}^{\ph}
\Gamma_2^{\ph}, \notag \\
\partial_+(\Gamma_3^{-1} \partial_- \Gamma_3^{\ph}) & = - C_{+3}^{\ph}
C_{-3}^{\ph} \Gamma_3^{-1} \Gamma_4^{\ph} + C_{+2}^{\ph} C_{-2}^{\ph}
\Gamma_2^{-1} \Gamma_3^{\ph}, \notag \\
& \vdots \label{e:2.34} \\
\partial_+(\Gamma_{s-1}^{-1} \partial_- \Gamma_{s-1}^{\ph}) & = -
C_{+(s-1)}^{\ph} C_{-(s-1)}^{\ph} \Gamma_{s-1}^{-1} \Gamma_s^{\ph} +
C_{+(s-2)}^{\ph} C_{-(s-2)}^{\ph} \Gamma_{s-2}^{-1} \Gamma_{s-1}
^{\ph}, \notag \\
\partial_+(\Gamma_s^{-1} \partial_- \Gamma_s^{\ph}) & = - C_{+s}^{\ph}
C_{-s}^{\ph} \Gamma_s^{-2} + C_{+(s-1)}^{\ph} C_{-(s-1)}^{\ph}
\Gamma_{s-1}^{-1} \Gamma_s^{\ph}. \notag
\end{align}
As above, in the case when the functions $C_{-\alpha}$ and
$C_{+\alpha}$ have no zeros, using the transformation (\ref{e:2.17})
and (\ref{e:2.18}), one can show that the equations (\ref{e:2.34}) are
equivalent to the same equations, where $C_{-\alpha} = C_-$ and
$C_{+\alpha} = C_+$ for some functions $C_-$ and $C_+$ which have no
zeros and are subject to the conditions (\ref{e:2.25}). If these
functions are real, then with the help of an appropriate change of the
coordinates $z^-$ and $z^+$ we can come to the equations
\begin{align*}
\partial_+(\Gamma_2^{-1} \partial_- \Gamma_2^{\ph}) & = - m^2
(\Gamma_2^{-1} \Gamma_3^{\ph} - \Gamma_2^{\ph}), \\
\partial_+(\Gamma_3^{-1} \partial_- \Gamma_3^{\ph}) & = - m^2 (
\Gamma_3^{-1} \Gamma_4^{\ph} - \Gamma_2^{-1} \Gamma_3^{\ph}), \\
& \vdots \\
\partial_+(\Gamma_{s-1}^{-1} \partial_- \Gamma_{s-1}^{\ph}) & = - m^2
(\Gamma_{s-1}^{-1} \Gamma_s^{\ph} - \Gamma_{s-2}^{-1} \Gamma_{s-1}
^{\ph}), \\
\partial_+(\Gamma_s^{-1} \partial_- \Gamma_s^{\ph}) & = - m^2
(\Gamma_s^{-2} - \Gamma_{s-1}^{-1} \Gamma_s^{\ph}),
\end{align*}
where $m$ is a nonzero constant, see also the papers
\cite{Mik81,MikOlsPer81}.

For $s = 2$ denoting $\Gamma_2$ by $\Gamma$ we have the equation
\[
\partial_+(\Gamma^{-1} \partial_- \Gamma^{\ph}) = - m^2 (\Gamma^{-2} -
\Gamma^{\ph}).
\]
Putting $\Gamma = \exp(F)$ we obtain
\[
\partial_+ \partial_- F = - m^2 [\exp(-2F) -  \exp(F)].
\]
This is the Tzitz\'eica equation \cite{Tzi08} which is now usually
called the Dodd--Bullough--Mikhailov equation \cite{DodBul77,Mik81}.

Let us show how the above equations are related to the Toda equations
associated with the $A_{2s-2}^{(2)}$ affine Lie group. Assume that $s
> 2$. Introduce an additional mapping $\Delta_0$ which satisfies the
equation
\begin{equation}
\partial_+(\Delta_0^{-1} \partial_- \Delta_0) = - \frac{m^2}{2}
\Gamma_2 \label{e:2.35}
\end{equation}
and denote
\[
\Delta_\alpha = 2^{-\alpha} \Delta_0^2 \prod_{\beta=2}^{\alpha+1}
\Gamma_\beta, \quad \alpha = 1, \ldots, s-2, \qquad \Delta_{s-1} =
2^{-s + 2} \Delta_0^2 \prod_{\beta=2}^s \Gamma_\beta.
\]
Now one can get convinced that the mappings $\Delta_\alpha$, $\alpha =
0, 1, \ldots, s-1$, satisfy the equations of the form (\ref{e:2.29})
where $n = s$ and $a_{\alpha \beta}$ are the matrix elements of the
Cartan matrix of an affine Lie algebra of type $A^{(2)}_{2s-2}$:
\[
(a_{\alpha \beta}) = \left( \begin{array}{rrrcrrr}
2 & -1 & 0 & \cdots & 0 & 0 & 0 \\
-2 & 2 & -1 & \cdots & 0 & 0 & 0 \\
0 & -1 & 2 & \cdots & 0 & 0 & 0 \\
\vdots & \vdots & \vdots & \ddots & \vdots & \vdots & \vdots \\
0 & 0 & 0 & \cdots & 2 & -1 & 0 \\
0 & 0 & 0 & \cdots & -1 & 2 & -1 \\
0 & 0 & 0 & \cdots & 0 & -2 & 2
\end{array} \right).
\]

In the case of $s=2$ we again define the mapping $\Delta_0$ with the
help of the relation (\ref{e:2.35}) and denote
\[
\Delta_1 = 2^{-1/3} \Delta_0^2 \Gamma_2.
\]
After an appropriate rescaling of the coordinates $z^-$ and $z^+$ we
come to the equations (\ref{e:2.29}) where $n = 2$ and $a_{\alpha
\beta}$ are the matrix elements of the Cartan matrix of an affine Lie
algebra of type $A^{(2)}_2$:
\[
(a_{\alpha \beta}) = \left( \begin{array}{rr}
2 & -1 \\
-4 & 2
\end{array} \right).
\]

\subsubsection{Third type}

In the case of an even $n = 2s$, $s \ge 2$, to come to an abelian Toda
system one should use again the automorphism $A$ defined by the
relation (\ref{e:2.31}) where now
\[
\psset{xunit=1.6em, yunit=1em}
B = \left( \raise -3.8\psyunit \hbox{\begin{pspicture}(-.3,.5)
(7.3,8.1)
\rput(1,8){$1$} \rput(0,7){$1$} \rput(7,6){$1$}
\qdisk(6.3,5.3){.7pt} \qdisk(6,5){.7pt} \qdisk(5.7,4.7){.7pt}
\rput(5,4){$1$} \rput(4,3){$-1$}
\qdisk(3.3,2.3){.7pt} \qdisk(3,2){.7pt} \qdisk(2.7,1.7){.7pt}
\rput(2,1){$-1$}
\psline(-.3,6.5)(7.3,6.5)
\psline(1.5,8.3)(1.5,.5)
\end{pspicture}} \right)
\]
and $h$ is a diagonal matrix with the diagonal matrix elements
\[
h_{11} = \epsilon_{2n-2}^{n-1} = \epsilon_{4s-2}^{2s-1} = -1, \qquad
h_{ii}= \epsilon_{2n-2}^{n-i+1} = \epsilon_{4s-2}^{2s-i+1}, \quad i =
2, \ldots, n.
\]
The number $p$ characterizing the block structure is equal to $n-1 =
2s-1$, $n_1 = 2$, and $n_\alpha = 1$ for $\alpha = 2, \ldots, 2s-1$.

The mapping $\Gamma_1$ is a mapping of $\bbR^2$ to the Lie group
$\rmSO_2(\bbC)$ which is isomorphic to $\bbC^\times$. Actually
$\Gamma_1$ is a $2 \times 2$ complex matrix valued function satisfying
the relation
\[
J_2^{-1} \: {}^{t\!} \Gamma_1^\ph \: J_2^\ph = \Gamma_1^{-1},
\]
where
\[
J_2 = \left( \begin{array}{cc}
0 & 1 \\ 1 & 0
\end{array} \right).
\]
It is easy to show that $\Gamma_1$ has the form
\[
\Gamma_1 = \left( \begin{array}{cc}
(\Gamma_1)_{11}^\ph & 0 \\ 0 & (\Gamma_1)_{11}^{-1}
\end{array} \right),
\]
where $(\Gamma_1)_{11}$ is a mapping of $\bbR^2$ to $\bbC^\times$. The
mappings $\Gamma_\alpha$, $\alpha = 2, \ldots, 2s-1$, are mappings of
$\bbR^2$ to $\bbC^\times$ satisfying the relations
\[
\Gamma_{2s-\alpha+1}^\ph = \Gamma^{-1}_\alpha.
\]

The mappings $C_{-1}$, $C_{+0}$ are complex $1 \times 2$ matrix valued
functions, the mappings $C_{-0}$, $C_{+1}$ are complex $2 \times 1$
matrix valued functions. Here one has
\begin{equation}
C_{-0} = J^{-1}_2 {}^{t\!} C_{-1}, \qquad C_{+0} = {}^{t\!} C_{+1}
J_2^\ph. \label{e:2.36}
\end{equation}
The mappings $C_{\pm \alpha}$, $\alpha = 2, \ldots, p - 1 = 2s - 2$,
are just complex functions, satisfying for $s > 2$ the equalities
\begin{equation}
C_{\pm(2s - \alpha)} = -C_{\pm \alpha}, \qquad \alpha = 2, \ldots,
s-1. \label{e:2.37}
\end{equation}

The mappings $(\Gamma_1)_{11}$ and $\Gamma_\alpha$, $\alpha = 2,
\ldots, s$, form a complete set of mappings parameterizing the mapping
$\gamma$. Taking into account the equalities (\ref{e:2.36}) and
(\ref{e:2.37}) we come to the following set of independent equations
equivalent to the Toda equation under consideration:
\begin{align*}
\partial_+((\Gamma_1^\ph)_{11}^{-1} \partial_- (\Gamma_1^\ph)_{11}
^\ph) = & - (C_{+1}^{\ph})_{11}^{\ph} (C_{-1}^{\ph})_{11}^{\ph}
(\Gamma_1^\ph)_{11}^{-1} \Gamma_2^{\ph} + (C_{+1}^{\ph})_{21}^\ph
(C_{-1}^{\ph})_{12}^\ph \Gamma_2^{\ph} (\Gamma_1^{\ph})_{11}^\ph, \\
\partial_+(\Gamma_2^{-1} \partial_- \Gamma_2^{\ph}) = & - C_{+2}^{\ph}
C_{-2}^{\ph} \Gamma_2^{-1} \Gamma_3^{\ph}
\\ & + (C_{+1}^{\ph})_{11}^{\ph} (C_{-1}^{\ph})_{11}^{\ph}
(\Gamma_1^\ph)_{11}^{-1} \Gamma_2^{\ph} + (C_{+1}^{\ph})_{21}^\ph
(C_{-1}^{\ph})_{12}^\ph \Gamma_2^{\ph} (\Gamma_1^{\ph})_{11}^\ph, \\
\partial_+(\Gamma_3^{-1} \partial_- \Gamma_3^{\ph}) = & - C_{+3}^{\ph}
C_{-3}^{\ph} \Gamma_3^{-1} \Gamma_4^{\ph} + C_{+2}^{\ph} C_{-2}^{\ph}
\Gamma_2^{-1} \Gamma_3^{\ph}, \\
& \vdots \\
\partial_+(\Gamma_{s-1}^{-1} \partial_- \Gamma_{s-1}^{\ph}) = & -
C_{+(s-1)}^{\ph} C_{-(s-1)}^{\ph} \Gamma_{s-1}^{-1} \Gamma_s^{\ph} +
C_{+(s-2)}^{\ph} C_{-(s-2)}^{\ph} \Gamma_{s-2}^{-1} \Gamma_{s-1}
^{\ph}, \\
\partial_+(\Gamma_s^{-1} \partial_- \Gamma_s^{\ph}) = & - C_{+s}^{\ph}
C_{-s}^{\ph} \Gamma_s^{-2} + C_{+(s-1)}^{\ph} C_{-(s-1)}^{\ph}
\Gamma_{s-1}^{-1} \Gamma_s^{\ph}.
\end{align*}
As well as for the first two types, under appropriate conditions these
equations can be reduced to the equations with $C_{-\alpha} = m$,
$C_{+\alpha} = m$ for $\alpha = 2, \ldots, s$, $(C_{-1})_{11} =
(C_{-1})_{12} = m / \sqrt{2}$ and $(C_{+1})_{11} = (C_{+1})_{21} = m /
\sqrt{2}$, where $m$ is a nonzero constant.\footnote{This choice is
convenient for applications of the rational dressing method.} Thus, we
come to the equations
\begin{align*}
\partial_+(\Gamma_1^{-1} \partial_- \Gamma_1^\ph) = & - \frac{m^2}{2}
(\Gamma_1^{-1} - \Gamma_1^\ph) \Gamma_2^{\ph}, \\
\partial_+(\Gamma_2^{-1} \partial_- \Gamma_2^{\ph}) = & - m^2
\Gamma_2^{-1} \Gamma_3^{\ph} + \frac{m^2}{2} (\Gamma_1^{-1} +
\Gamma_1^{\ph})\Gamma_2^{\ph}, \\
\partial_+(\Gamma_3^{-1} \partial_- \Gamma_3^{\ph}) = & - m^2
(\Gamma_3^{-1} \Gamma_4^{\ph} - \Gamma_2^{-1} \Gamma_3^{\ph}), \\
& \vdots \\
\partial_+(\Gamma_{s-1}^{-1} \partial_- \Gamma_{s-1}^{\ph}) = & - m^2
(\Gamma_{s-1}^{-1} \Gamma_s^{\ph} - \Gamma_{s-2}^{-1} \Gamma_{s-1}
^{\ph}), \\
\partial_+(\Gamma_s^{-1} \partial_- \Gamma_s^{\ph}) = & - m^2
(\Gamma_s^{-2} - \Gamma_{s-1}^{-1} \Gamma_s^{\ph}),
\end{align*}
where slightly abusing notation we denote $(\Gamma_1)_{11}$ by
$\Gamma_1$.

Introduce now an additional mapping $\Delta_0$ which satisfies the
equation
\[
\partial_+(\Delta_0^{-1} \partial_- \Delta_0) = - \frac{m^2}{2}
\Gamma_1 \Gamma_2
\]
and denote
\[
\Delta_1 = \Delta_0 \Gamma_1, \qquad \Delta_\alpha =
2^{\alpha(\alpha-1)/(2s-1) - \alpha + 1}\Delta_0^2 \prod_{\beta=1}
^\alpha \Gamma_\beta, \quad \alpha = 2, \ldots, s.
\]
The mappings $\Delta_\alpha$, $\alpha = 0, 1, \ldots, s$, satisfy the
equations which, after an appropriate rescaling of the coordinates
$z^-$ and $z^+$, take the form (\ref{e:2.29}), where now $n = s + 1$
and $a_{\alpha \beta}$ are the matrix elements of the Cartan matrix of
an affine Lie algebra of type $A^{(2)}_{2s-1}$:
\[
(a_{\alpha \beta}) = \left( \begin{array}{rrrcrrr}
2 & 0 & -1 & \cdots & 0 & 0 & 0 \\
0 & 2 & -1 & \cdots & 0 & 0 & 0 \\
-1 & -1 & 2 & \cdots & 0 & 0 & 0 \\
\vdots & \vdots & \vdots & \ddots & \vdots & \vdots & \vdots \\
0 & 0 & 0 & \cdots & 2 & -1 & 0 \\
0 & 0 & 0 & \cdots & -1 & 2 & -1 \\
0 & 0 & 0 & \cdots & 0 & -2 & 2
\end{array} \right).
\]

\section{Soliton solutions}

In this section we compare two methods used to construct soliton
solutions of the abelian Toda systems associated with the loop groups
of the complex general linear groups. We restrict ourselves to the
abelian Toda equations of the first type which have the form
(\ref{e:2.27}).

\subsection{Hirota's method} \label{s:3.1}

It is convenient to treat the system (\ref{e:2.27}) as an infinite
system
\begin{equation}
\partial_+ (\Gamma^{-1}_\alpha \partial_- \Gamma^{}_\alpha) = - m^2 (
\Gamma^{-1}_\alpha \Gamma^{}_{\alpha+1} - \Gamma^{-1}_{\alpha-1}
\Gamma^{}_\alpha), \label{e:3.1}
\end{equation}
where the functions $\Gamma_\alpha$ are defined for arbitrary integer
values of the index $\alpha$ in the periodic way,
\begin{equation}
\Gamma_{\alpha+n} = \Gamma_\alpha. \label{e:3.2}
\end{equation}

Following the Hirota's approach \cite{Hol92, ConFerGomZim93, MacMcG92,
AraConFerGomZim93, ZhuCal93}, one introduces $\tau$-functions
connected with $\Gamma_\alpha$ by the relation
\begin{equation}
\Gamma_\alpha = \tau_\alpha/\tau_{\alpha-1}, \label{e:3.3}
\end{equation}
where we assume that the $\tau$-functions are defined for all integer
values of the index $\alpha$. This change of variables is the essence
of the Hirota's method. The periodicity condition (\ref{e:3.2}) in
terms of the $\tau$-functions takes the form
\[
\tau_{\alpha+n}/\tau_\alpha = \tau_{\alpha-1+n}/\tau_{\alpha-1}.
\]
It means that the ratio $\tau_{\alpha+n}/\tau_\alpha$ does not depend
on $\alpha$.
Noting that
\[
\Gamma = \Gamma_1 \Gamma_2 \ldots \Gamma_n = \tau_n/\tau_0,
\]
we write
\[
\tau_{\alpha+n} = \Gamma \tau_\alpha.
\]
As was explained in the previous section, with the appropriate
symmetry
transformation one can make $\Gamma = 1$. We will assume that the
corresponding symmetry transformation was performed, and, therefore,
\begin{equation}
\tau_{\alpha+n} = \tau_\alpha. \label{e:3.4}
\end{equation}

The equations (\ref{e:3.1}) in terms of the $\tau$-functions look as
\[
\partial_+ (\tau_\alpha^{-1} \, \partial_- \tau_\alpha^{}) -
\partial_+ (\tau_{\alpha-1}^{-1} \, \partial_- \tau_{\alpha-1}^{}) = -
m^2 (\tau_{\alpha-1}^{} \, \tau_\alpha^{-2} \, \tau_{\alpha+1}^{} -
\tau_{\alpha-2}^{} \, \tau_{\alpha-1}^{-2} \, \tau_\alpha^{}).
\]
Consider the following decoupling of the above equations
\begin{equation}
\partial_+ (\tau_\alpha^{-1} \, \partial_- \tau_\alpha^{}) = m^2 \, (1
- \tau_{\alpha-1}^{} \, \tau_\alpha^{-2} \, \tau_{\alpha+1}^{}).
\label{e:3.5}
\end{equation}
It is evident that if the $\tau$-functions satisfy these equations,
then the functions $\Gamma_\alpha$ defined by (\ref{e:3.3}) satisfy
the system (\ref{e:2.27}). Moreover, it is easy to show that in this
case the functions
\[
\Delta_\alpha = \exp(-m^2 z^+ z^-) \tau_\alpha
\]
satisfy the system (\ref{e:2.28}).

It is convenient to rewrite the equations (\ref{e:3.5}) in the form
\begin{equation}
\tau_\alpha^{} \, \partial_+ \partial_- \tau_\alpha - \partial_+
\tau_\alpha \, \partial_- \tau_\alpha = m^2 \, (\tau_\alpha^2 -
\tau_{\alpha-1}^{} \, \tau_{\alpha+1}^{}). \label{e:3.6}
\end{equation}
These equations are of the Hirota bilinear type. Their solutions,
leading to multi-soliton solutions of the system (\ref{e:2.27}), can
be found perturbatively in the following way.

Consider a series expansion of the functions $\tau_\alpha$ in some
parameter $\varepsilon$ which will be set to one at the final step of
the construction. So we represent the functions $\tau_\alpha$ in the
form
\begin{equation}
\tau_\alpha^{} = \tau_\alpha^{(0)} + \varepsilon \tau_\alpha^{(1)} +
\varepsilon^2 \tau_\alpha^{(2)} + \ldots, \label{e:3.7}
\end{equation}
and assume that $\tau_\alpha^{(0)}$ are constants. The periodicity
condition
(\ref{e:3.4}) gives
\[
\tau_{\alpha+n}^{(k)} = \tau_\alpha^{(k)}, \qquad k = 0, 1, \ldots.
\]

Let us try to solve equations (\ref{e:3.6}) order by order in
$\varepsilon$. Actually our goal is to find solutions for which the
series (\ref{e:3.7}) truncates at some finite order in $\varepsilon$.
In such a case we have an exact solution.

Using the expansion (\ref{e:3.7}), one obtains
\[
\tau_\alpha \, \partial_+ \partial_- \tau_\alpha = \sum_{k=0}^\infty
\varepsilon^k \sum_{\ell=0}^k \tau^{(k-\ell)}_\alpha \partial_+
\partial_- \tau^{(\ell)}_\alpha.
\]
Similarly one has
\[
\partial_+ \tau_\alpha \, \partial_- \tau_\alpha = \sum_{k=0}^\infty
\varepsilon^k \sum_{\ell=0}^k \partial_+ \tau^{(k-\ell)}_\alpha
\partial_- \tau^{(\ell)}_\alpha.
\]
Now, using the equality
\[
\tau_\alpha^2 - \tau_{\alpha-1}^{} \, \tau_{\alpha+1}^{} = \sum_{k=0}
^\infty \varepsilon^k \sum_{\ell=0}^k \left(\tau_\alpha^{(\ell)} \,
\tau_\alpha^{(k-\ell)} - \tau_{\alpha-1}^{(\ell)} \, \tau_{\alpha+1}
^{(k-\ell)} \right),
\]
we see that the equations (\ref{e:3.6}) are equivalent to the
equations
\begin{equation}
\sum_{\ell=0}^k \left( \tau_\alpha^{(k-\ell)} \partial_+ \partial_-
\tau_\alpha^{(\ell)} - \partial_+ \tau_\alpha^{(k-\ell)} \partial_-
\tau_\alpha^{(\ell)} \right) = m^2 \sum_{\ell=0}^k \left(
\tau_\alpha^{(\ell)} \, \tau_\alpha^{(k-\ell)} - \tau_{\alpha-1}
^{(\ell)} \, \tau_{\alpha+1}^{(k-\ell)} \right), \label{e:3.8}
\end{equation}
which can be solved step by step starting from $k = 0$.

For $k=0$ one has
\[
\tau_{\alpha-1}^{(0)} \, \tau_{\alpha+1}^{(0)} - \tau_\alpha^{(0)} \,
\tau_\alpha^{(0)} = 0,
\]
that can be rewritten as
\[
\tau_{\alpha+1}^{(0)}/\tau_\alpha^{(0)} = \tau_\alpha^{(0)}
/\tau_{\alpha-1}^{(0)}.
\]
It is clear that the general solution to this relation is
\begin{equation}
\tau_\alpha^{(0)} = \tau_0^{(0)} d^\alpha, \label{e:3.9}
\end{equation}
where $d$ is an arbitrary constant. Recall that the Toda equations
(\ref{e:3.1}) are invariant with respect to the multiplication of all
$\Gamma_\alpha$ by the same constant. From the point of view of the
$\tau$-functions this is equivalent to the multiplication of the
function $\tau_\alpha$ by the $\alpha$th power of the constant. Hence,
different values of the constant $d$ in the relation (\ref{e:3.9})
correspond to the functions $\Gamma_\alpha$ connected by a rescaling.
Moreover, dividing all $\tau$-functions by the same constant we do not
change the functions $\Gamma_\alpha$. Therefore, actually without any
loss of generality, one can put
\begin{equation}
\tau_\alpha^{(0)} = 1. \label{e:3.10}
\end{equation}
Using this equality, we rewrite (\ref{e:3.8}) as
\begin{multline}
\partial_+ \partial_- \tau^{(k)}_\alpha - m^2 \sum_{\beta=0}^{n-1}
a^{}_{\alpha \beta} \, \tau^{(k)}_\beta = - \sum_{\ell=1}^{k-1} \bigl(
\tau_\alpha^{(k-\ell)} \partial_+ \partial_-\tau_\alpha^{(\ell)} -
\partial_+ \tau_\alpha^{(k-\ell)} \partial_- \tau_\alpha^{(\ell)}
\bigr) \\
+ m^2 \sum_{\ell=1}^{k-1} \left( \tau_\alpha^{(\ell)} \,
\tau_\alpha^{(k-\ell)} - \tau_{\alpha-1}^{(\ell)} \, \tau_{\alpha+1}
^{(k-\ell)} \right), \label{e:3.11}
\end{multline}
where $a_{\alpha \beta}$ are the matrix elements of the Cartan matrix
(\ref{e:2.30}) of an affine Lie algebra of type $A^{(1)}_{n-1}$. Thus,
we see that at each step we should solve a system of linear
differential equations.

In particular, for $k=1$ one has to solve the system of equations
\begin{equation}
\partial_+ \partial_- \tau^{(1)}_\alpha - m^2 \sum_{\beta=0}^{n-1}
a^{}_{\alpha \beta} \, \tau^{(1)}_\beta = 0. \label{e:3.12}
\end{equation}
It is easy to find solutions of these equations using the eigenvectors
$\theta_\rho$ of the Cartan matrix $(a_{\alpha \beta})$ which are
given by
\begin{equation}
(\theta_\rho)_\alpha = \epsilon_n^{(\alpha + 1) \rho}, \qquad \rho =
0, 1, \ldots, n-1. \label{e:3.13}
\end{equation}
Here the corresponding eigenvalues are
\[
\kappa_\rho^2 = 2 - \epsilon_n^\rho - \epsilon_n^{-\rho} = 4 \sin^2 (
\pi \rho/n ).
\]

Let us assume that the functions $\tau_\alpha^{(1)}$ are of the form
\begin{equation}
\tau_\alpha^{(1)} = \sum_{i = 1}^r E_{\alpha i}, \label{e:3.14}
\end{equation}
where
\begin{equation}
E_{\alpha i} = \epsilon_n^{(\alpha + 1) \rho_i} \exp[ m \,
\kappa_{\rho_i} (\zeta_i^{-1} z^- + \zeta_i^{} z^+) + \delta_i ].
\label{e:3.15}
\end{equation}
Here $\rho_i$ is an integer from the interval from $1$ to $n-1$,
$\zeta_i$ and $\delta_i$ are arbitrary complex numbers. Note that the
choice $\rho_i = 0$ is excluded because it gives a constant
contribution to the $\tau$-functions which can be included into
$\tau_\alpha^{(0)}$. Then after a corresponding rescaling one can
satisfy the normalization (\ref{e:3.10}). For definiteness we assume
that
\begin{equation}
\kappa_\rho = - \rmi (\epsilon_n^{\rho/2} - \epsilon_n^{-\rho/2}) = 2
\sin ( \pi \rho/n ).
\label{e:3.16}
\end{equation}
Certainly, the ansatz (\ref{e:3.14}) does not give a general solution
to the equations (\ref{e:3.12}) but it ensures truncation of the
expansion (\ref{e:3.7}).

For $k=2$ the equations (\ref{e:3.11}) have the form
\begin{multline*}
\partial_+ \partial_- \tau_\alpha^{(2)} - m^2 \sum_{\beta = 0}^{n-1}
a_{\alpha \beta} \tau_\beta^{(2)} \\
= - \tau_\alpha^{(1)} \partial_+ \partial_- \tau_\alpha^{(1)} +
\partial_+ \tau_\alpha^{(1)} \partial_- \tau_\alpha^{(1)} + m^2
(\tau_\alpha^{(1)} \tau_\alpha^{(1)} - \tau_{\alpha-1}^{(1)}
\tau_{\alpha+1}^{(1)}).
\end{multline*}
Using the equalities
\[
\partial_- E_{\alpha i} = m \, \kappa_{\rho_i} \zeta^{-1}_i E_{\alpha
i}, \qquad \partial_+ E_{\alpha i} = m \, \kappa_{\rho_i} \zeta_i \,
E_{\alpha i},
\]
one obtains
\begin{multline}
- \tau_\alpha^{(1)} \partial_+ \partial_- \tau_\alpha^{(1)} +
\partial_+ \tau_\alpha^{(1)} \partial_- \tau_\alpha^{(1)} + m^2
(\tau_\alpha^{(1)} \tau_\alpha^{(1)} - \tau_{\alpha-1}^{(1)}
\tau_{\alpha+1}^{(1)}) \\
= \frac{m^2}{2} \sum_{i_1, i_2 = 1}^r \left[\kappa_{\rho_{i_1}}
\kappa_{\rho_{i_2}} \left( \zeta_{i_1}^{} \zeta_{i_2}^{-1} +
\zeta_{i_1}^{-1} \zeta_{i_2}^{} \right) - \kappa_{\rho_{i_1}}^2 -
\kappa_{\rho_{i_2}}^2 + \kappa_{\rho_{i_1} - \rho_{i_2}}^2 \right]
E_{\alpha i_1} E_{\alpha i_2}. \label{e:3.17}
\end{multline}

Note that in the case of $r = 1$ we have
\[
- \tau_\alpha^{(1)} \partial_+ \partial_- \tau_\alpha^{(1)} +
\partial_+ \tau_\alpha^{(1)} \partial_- \tau_\alpha^{(1)} + m^2
(\tau_\alpha^{(1)} \tau_\alpha^{(1)} - \tau_{\alpha-1}^{(1)}
\tau_{\alpha+1}^{(1)}) = 0
\]
and we can put $\tau_\alpha^{(2)} = 0$. This gives the one-soliton
solutions
\begin{equation}
\Gamma_\alpha = \frac{1 + \epsilon_n^{\rho(\alpha+1)} \exp[m \,
\kappa_\rho (\zeta^{-1} z^-
+ \zeta z^+) + \delta]}{1 + \epsilon_n^{\rho \alpha} \exp[m \,
\kappa_\rho (\zeta^{-1}
z^- + \zeta z^+) + \delta]}. \label{e:3.18}
\end{equation}

In the case when $r > 1$ the equality (\ref{e:3.17}) suggests to look
for $\tau_\alpha^{(2)}$ of the form
\[
\tau_\alpha^{(2)} = \frac{1}{2} \sum_{i_1, i_2 = 1}^r \eta_{i_1 i_2}
E_{\alpha i_1} E_{\alpha i_2}.
\]
Here one can easily find that
\begin{multline*}
\partial_+ \partial_- \tau_\alpha^{(2)} - m^2 \sum_{\beta = 0}^{n - 1}
a_{\alpha \beta} \tau_\beta^{(2)} \\
= \frac{m^2}{2} \sum_{i_1, i_2 = 1}^r \left[\kappa_{\rho_{i_1}}
\kappa_{\rho_{i_2}} ( \zeta_{i_1}^{} \zeta_{i_2}^{-1} +\zeta_{i_1}
^{-1} \zeta_{i_2} ) +
\kappa_{\rho_{i_1}}^2 + \kappa_{\rho_{i_2}}^2 - \kappa_{\rho_{i_1} +
\rho_{i_2}}^2 \right] \eta_{i_1 i_2} E_{\alpha i_1} E_{\alpha i_2}
\end{multline*}
and, therefore,
\[
\eta_{i_1 i_2} = \frac{\displaystyle \kappa_{\rho_{i_1}}
\kappa_{\rho_{i_2}} ( \zeta_{i_1}^{} \zeta_{i_2}^{-1} +\zeta_{i_1}
^{-1} \zeta_{i_2} ) - \kappa_{\rho_{i_1}}^2 - \kappa_{\rho_{i_2}}^2 +
\kappa_{\rho_{i_1} - \rho_{i_2}}^2}{\kappa_{\rho_{i_1}}
\kappa_{\rho_{i_2}} ( \zeta_{i_1}^{} \zeta_{i_2}^{-1} +\zeta_{i_1}
^{-1} \zeta_{i_2} ) +
\kappa_{\rho_{i_1}}^2 + \kappa_{\rho_{i_2}}^2 - \kappa_{\rho_{i_1} +
\rho_{i_2}}^2},
\]
that can be written as
\begin{equation}
\eta_{i_1 i_2}= \frac{( \zeta_{i_1}^{} \zeta_{i_2}^{-1} +\zeta_{i_1}
^{-1} \zeta_{i_2} ) - 2 \cos [ \pi (\rho_{i_1} - \rho_{i_2})/n ]}{(
\zeta_{i_1}^{} \zeta_{i_2}^{-1} +\zeta_{i_1}^{-1} \zeta_{i_2} ) - 2
\cos [ \pi (\rho_{i_1} + \rho_{i_2})/n ]}. \label{e:3.19}
\end{equation}
The quantities $\eta_{i_1 i_2}$ are symmetric with respect to the
indices $i_1$, $i_2$ and they turn to zero when $i_1 = i_2$. Hence one
can write
\[
\tau_\alpha^{(2)} = \sum_{1 \le i_1 < i_2 \le r} \eta_{i_1 i_2}
E_{\alpha i_1} E_{\alpha i_2}.
\]
It can be shown that when $r = 2$ one can choose $\tau_\alpha^{(3)} =
0$. In general, it can be shown that for $\ell \le r$ one can choose
\[
\tau_\alpha^{(\ell)} = \sum_{1 \le i_1 < i_2 < \ldots < i_\ell \le r}
\left(\prod_{1 \le j < k \le \ell} \eta_{i_j i_k} \right) E_{\alpha
i_1} E_{\alpha i_2} \ldots
E_{\alpha i_\ell}
\]
and $\tau_i^{(\ell)} = 0$ for $\ell > r$. In other words, the
equations (\ref{e:3.6}) have the following solutions
\begin{equation}
\tau_\alpha = 1 + \sum_{i = 1}^r E_{\alpha i} + \sum_{\ell = 2}^r
\left[ \sum_{1 \le i_1 < i_2 < \ldots < i_\ell \le r} \left(\prod_{1
\le j < k \le \ell} \eta_{i_j i_k} \right) E_{\alpha i_1} E_{\alpha
i_2} \ldots E_{\alpha i_\ell} \right]. \label{e:3.20}
\end{equation}

\subsection{Rational dressing} \label{s:3.2}

Since for any $\bar{m} \in \bbR^2$ the matrices $c_-(\bar{m})$ and 
$c_+(\bar{m})$
commute,\footnote{Actually in the case under consideration $c_-$ and
$c_+$ are constant mappings.} it is obvious that
\begin{equation}
\gamma = I_n,
\label{e:3.21}
\end{equation}
where $I_n$ is the $n \times n$ unit matrix, is a solution to the Toda
equation (\ref{e:2.15}). Denote a mapping of $\bbR^2 \times S^1$ to
$\rmGL_n(\bbC)$ which generates the corresponding connection by
$\varphi$. Using the equalities (\ref{e:2.11}) and (\ref{e:2.14}) and
remembering that in our case $L = 1$, we write
\[
\varphi^{-1} \partial_- \varphi = \lambda^{-1} c_-, \qquad
\varphi^{-1} \partial_+ \varphi = \lambda \: c_+,
\]
where the matrices $c_-$ and $c_+$ are defined by the relation
(\ref{e:2.26}).

To construct more interesting solutions to the Toda equations we will
look for a mapping $\psi$, such that the mapping
\begin{equation}
\varphi' = \varphi \: \psi
\label{e:3.22}
\end{equation}
would generate a connection satisfying the grading condition and the
gauge-fixing constraint $\omega_{+0} = 0$.

For any $\bar{m} \in \bbR^2$ the mapping $\tilde \psi_m$ defined by the
equality $\tilde \psi_m(\bar{p}) = \psi(\bar{m}, \bar{p})$, 
$\bar{p} \in S^1$, is a smooth mapping of $S^1$ to $\rmGL_n(\bbC)$. 
Recall that we treat $S^1$ as a subset of the complex plane which, 
in turn, will be treated as a subset of the Riemann sphere. Assume 
that it is possible to extend analytically each mapping $\tilde \psi_m$ 
to all of the Riemann sphere. As the result we get a mapping of the 
direct product of $\bbR^2$ and the Riemann sphere to $\rmGL_n(\bbC)$ 
which we also denote by $\psi$. Suppose that for any $\bar{m} \in \bbR^2$ 
the analytic extension of $\tilde \psi_m$ results in a rational mapping 
regular at the points $0$ and $\infty$, hence the name rational dressing. 
Below, for each point $\bar{p}$ of the Riemann sphere we denote by $\psi_p$ 
the mapping of $\bbR^2$ to $\rmGL_n(\bbC)$ defined by the equality
$\psi_p(\bar{m}) = \psi(\bar{m}, \bar{p})$.

Since we deal with the Toda equations described in section \ref{s:2.3.1},
that is, the mapping $\psi$ is generated by a mapping of $\bbR^2$ to the 
loop group $\calL_{a, n}(\rmGL_n(\bbC))$ with the automorphism $a$ defined 
by the relations (\ref{e:aut}) and (\ref{e:2.22}), for any 
$\bar{m} \in \bbR^2$ and $\bar{p} \in S^1$ we should have
\begin{equation}
\psi(\bar{m}, \epsilon_n \bar{p}) 
= h \: \psi(\bar{m}, \bar{p}) \: h^{-1},
\label{e:3.23}
\end{equation}
where $h$ is a diagonal matrix described by the relation
(\ref{e:2.22}). The equality (\ref{e:3.23}) means that two rational
mappings coincide on $S^1$, therefore, they must coincide on the
entire Riemann sphere.

A mapping, satisfying the equality (\ref{e:3.23}), can be constructed
by the following procedure. Let $\chi$ be an arbitrary mapping of the
direct product of $\bbR^2$ and the Riemann sphere to $\rmGL_n(\bbC)$.
Let $\hat a$ be a linear operator acting on $\chi$ as
\[
\hat a \: \chi (\bar{m}, \bar{p}) 
= h \: \chi(\bar{m}, \epsilon_n^{-1} \bar{p}) \: h^{-1}.
\]
It is easy to get convinced that the mapping
\[
\psi = \sum_{k=1}^n \hat a^k \chi
\]
satisfies the relation $\hat a \: \psi = \psi$ which is equivalent to
the equality (\ref{e:3.23}). Note that $\hat a^n \chi = \chi$.

To construct a rational mapping satisfying (\ref{e:3.23}) we start
with a rational mapping regular at the points $0$ and $\infty$ and
having poles at $r$ different nonzero points $\mu_i$, $i = 1, \ldots,
r$. Concretely speaking, we consider a mapping $\chi$ of the form
\[
\chi = \left( I_n + n \sum_{i=1}^r \frac{\lambda}{\lambda -
\mu_i} P_i \right) \chi_0,
\]
where $P_i$ are some smooth mappings of $\bbR^2$ to the algebra
$\Mat_n(\bbC)$ of $n \times n$ complex matrices and $\chi_0$ is a
mapping of $\bbR^2$ to the Lie subgroup of $\rmGL_n(\bbC)$ formed by
the elements $g \in \rmGL_n(\bbC)$ satisfying the equality
\begin{equation}
h g h^{-1} = g. \label{e:3.24}
\end{equation}
Actually this subgroup coincides with the subgroup $G_0$. The
averaging procedure leads to the mapping
\begin{equation}
\psi = \left(I_n + \sum_{i=1}^r \sum_{k=1}^n \frac{\lambda}
{\lambda - \epsilon_n^k \mu_i} h^k P_i \: h^{-k} \right) \psi_0,
\label{e:3.25}
\end{equation}
where $\psi_0 = n \chi_0$. It is convenient to assume that $\mu_i^n
\ne \mu_j^n$ for all $i \ne j$.

Denote by $\psi^{-1}$ the mapping of $\bbR^2 \times S^1$ to
$\rmGL_n(\bbC)$ defined by the relation
\[
\psi^{-1}(\bar{m}, \bar{p}) = (\psi(\bar{m}, \bar{p}))^{-1}.
\]
Suppose that for any fixed $\bar{m} \in \bbR^2$ the mapping
$\tilde\psi_m^{-1}$ of $S^1$ to $\rmGL_n(\bbC)$ can be extended
analytically to a mapping of the Riemann sphere to $\rmGL_n(\bbC)$ and
as the result we obtain a rational mapping of the same structure as
the mapping $\psi$,
\begin{equation}
\psi^{-1} = \psi_0^{-1} \left( I_n 
+ \sum_{i = 1}^r \sum_{k=1}^n \frac{\lambda}
{\lambda - \epsilon_n^k \nu_i} h^k Q_i h^{-k} \right),
\label{e:3.26}
\end{equation}
with the pole positions satisfying the conditions $\nu_i \ne 0$,
$\nu_i^n \ne \nu_j^n$ for all $i \ne j$, and additionally $\nu_i^n \ne
\mu_j^n$ for any $i$ and $j$. We will denote the mapping of the direct
product of $\bbR^2$ and the Riemann sphere to $\rmGL_n(\bbC)$ again by
$\psi^{-1}$.

By definition, the equality
\[
\psi^{-1} \psi = I_n
\]
is valid at all points of the direct product of $\bbR^2$ and $S^1$.
Since $\psi^{-1} \psi$ is a rational mapping, the above equality is
valid at all points of the direct product of $\bbR^2$ and the Riemann
sphere. Hence, the residues of $\psi^{-1} \psi$ at the points $\nu_i$
and $\mu_i$ should be equal to zero. Explicitly we have
\begin{gather}
Q_i \left( I_n + \sum_{j = 1}^r \sum_{k=1}^n \frac{\nu_i}{\nu_i -
\epsilon_n^k \mu_j} h^k P_j h^{-k} \right) = 0,
\label{e:3.27} \\
\left( I_n + \sum_{j = 1}^r \sum_{k=1}^n \frac{\mu_i}{\mu_i -
\epsilon_n^k \nu_j} h^k Q_j h^{-k}
\right) P_i = 0.
\label{e:3.28}
\end{gather}
In this case due to the relation
(\ref{e:3.23}) the residues of $\psi^{-1} \psi$ at the points
$\epsilon^k_n \mu_i$ and $\epsilon^k_n \nu_i$ vanish for $k = 1, \ldots, n$.
We will discuss later how to satisfy these relations, and now let us
consider what connection is generated by the mapping $\varphi'$
defined by (\ref{e:3.22}) with the mapping $\psi$ possessing the
prescribed properties.

Using the representation (\ref{e:3.22}), we obtain for the components
of the connection generated by $\varphi'$ the expressions
\begin{gather}
\omega_- = \psi^{-1} \partial_- \psi + \lambda^{-1} \psi^{-1} c_-
\psi, \label{e:3.29} \\
\omega_+ = \psi^{-1} \partial_+ \psi + \lambda \psi^{-1} c_+ \psi.
\label{e:3.30}
\end{gather}
We see that the component $\omega_-$ is a rational mapping which has
simple poles at the points $\mu_i$, $\nu_i$ and zero.\footnote{Here
and below discussing the holomorphic properties of mappings and
functions we assume that the point of the space $\bbR^2$ is arbitrary
but fixed.} Similarly, the component $\omega_+$ is a rational mapping
which has simple poles at the points $\mu_i$, $\nu_i$ and infinity. We
are looking for a connection which satisfies the grading and gauge-
fixing conditions. The grading condition in our case is the
requirement that for each point of $\bbR^2$ the component $\omega_-$
is rational and has the only simple pole at zero, while the component
$\omega_+$ is rational and has the only simple pole at infinity.
Hence, we demand that the residues of $\omega_-$ and $\omega_+$ at the
points $\mu_i$ and $\nu_i$ should vanish. In this case, as above, due 
to the relation (\ref{e:3.23}) the residues of $\omega_-$ and $\omega_+$ 
at the points $\epsilon^k_n \mu_i$ and $\epsilon^k_n \nu_i$ vanish for 
$k = 1, \ldots, n$.

The residues of $\omega_-$ and $\omega_+$ at the points $\nu_i$ are
equal to zero if and only if
\begin{gather}
(\partial_- Q_i - \nu_i^{-1} Q_i c_-) \left( I_n + \sum_{j = 1}^r
\sum_{k=1}^n \frac{\nu_i}{\nu_i - \epsilon_n^k \mu_j} h^k P_j h^{-k}
\right) = 0,
\label{e:3.31} \\
(\partial_+ Q_i - \nu_i Q_i c_+) \left( I_n + \sum_{j = 1}^r \sum_{k=
1}^n \frac{\nu_i}{\nu_i - \epsilon_n^k \mu_j} h^k P_j h^{-k} \right) =
0,
\label{e:3.32}
\end{gather}
respectively. Similarly, the requirement of vanishing of the residues
at the points $\mu_i$ gives the relations
\begin{gather}
\left( I_n + \sum_{j = 1}^r \sum_{k=1}^n \frac{\mu_i}{\mu_i -
\epsilon_n^k \nu_j} h^k Q_j h^{-k} \right) (\partial_- P_i +
\mu_i^{-1} c_- P_i) = 0,
\label{e:3.33} \\
\left( I_n + \sum_{j = 1}^r \sum_{k=1}^n \frac{\mu_i}{\mu_i -
\epsilon_n^k \nu_j} h^k Q_j h^{-k} \right) (\partial_+ P_i + \mu_i c_+
P_i) = 0.
\label{e:3.34}
\end{gather}
To obtain the relations (\ref{e:3.31})--(\ref{e:3.34}) we made 
use of the equalities (\ref{e:3.27}), (\ref{e:3.28}).

Suppose that we have succeeded in satisfying the relations
(\ref{e:3.27}), (\ref{e:3.28}) and (\ref{e:3.31})--(\ref{e:3.34}). In
such a case from the equalities (\ref{e:3.29}) and (\ref{e:3.30}) it
follows that the connection under consideration satisfies the grading
condition.

It is easy to see from (\ref{e:3.30}) that
\[
\omega_+(\bar{m}, 0) = \psi_0^{-1}(\bar{m}) 
\partial_+ \psi_0(\bar{m}).
\]
Taking into account that $\omega_{+0}(\bar{m}) = \omega_+(\bar{m}, 0)$, 
we conclude that the gauge-fixing constraint $\omega_{+0} = 0$ is
equivalent to the relation
\begin{equation}
\partial_+ \psi_0 = 0. \label{e:3.35}
\end{equation}
Assuming that this relation is satisfied, we come to a connection
satisfying both the grading condition and the gauge-fixing condition.

Recall that if a flat connection $\omega$ satisfies the grading and
gauge-fixing conditions, then there exist a mapping $\gamma$ from
$\bbR^2$ to $G$ and mappings $c_-$ and $c_+$ of $\bbR^2$ to
$\gothg_{-1}$ and $\gothg_{+1}$, respectively, such that the
representation (\ref{e:2.14}) for the components $\omega_-$  and
$\omega_+$ is valid. In general, the mappings $c_-$ and $c_+$
parameterizing the connection components may be different from the
mappings $c_-$ and $c_+$ which determine the mapping $\varphi$. Let us
denote the mappings corresponding to the connection under
consideration by $\gamma'$, $c_-'$ and $c_+'$. Thus, we have
\begin{align}
\psi^{-1} \partial_- \psi + \lambda^{-1} \psi^{-1} c_- \psi &=
\gamma^{\prime -1} \partial_- \gamma' + \lambda^{-1} c_-',
\label{e:3.36} \\
\psi^{-1} \partial_+ \psi + \lambda \psi^{-1} c_+ \psi &= \lambda
\gamma^{\prime -1} c_+' \gamma'. \label{e:3.37}
\end{align}
Note that $\psi_\infty$ is a mapping of $\bbR^2$ to the Lie subgroup
of $\rmGL_n(\bbC)$ defined by the relation (\ref{e:3.24}). Recall that
this subgroup coincides with $G_0$, and denote $\psi_\infty$ by
$\gamma$. From the relation (\ref{e:3.36}) we obtain the equality
\[
\gamma^{\prime -1} \partial_- \gamma' = \gamma^{-1} \partial_- \gamma.
\]
The same relation (\ref{e:3.36}) gives
\[
\psi^{-1}_0 c_- \psi_0 = c_-'.
\]
Impose the condition $\psi_0 = I_n$, which is consistent with
(\ref{e:3.35}). Here we have
\[
c_-' = c_-.
\]
Finally, from (\ref{e:3.37}) we obtain
\[
\gamma^{\prime -1} c_+' \gamma' = \gamma^{-1} c_+ \gamma.
\]
We see that if we impose the condition $\psi_0 = I_n,$ then the
components of the connection under consideration have the form given
by (\ref{e:2.14}) where $\gamma = \psi_\infty$.

Thus, to find solutions to Toda equations under consideration, we can
use the following procedure. Fix $2r$ complex numbers $\mu_i$ and
$\nu_i$. Find matrix-valued functions $P_i$ and $Q_i$ satisfying the
relations (\ref{e:3.27}), (\ref{e:3.28}) and
(\ref{e:3.31})--(\ref{e:3.34}). With the help of (\ref{e:3.25}),
(\ref{e:3.26}), assuming that
\[
\psi_0 = I_n,
\]
construct the mappings $\psi$ and $\psi^{-1}$. Then, the mapping
\begin{equation}
\gamma = \psi_\infty
\label{e:3.38}
\end{equation}
satisfies the Toda equation (\ref{e:2.15}).

Let us return to the relations (\ref{e:3.27}), (\ref{e:3.28}). It can
be shown that, if we suppose that the matrices $P_i$ and $Q_i$ are of
maximum rank, then we get the trivial solution of the Toda equation
given by (\ref{e:3.21}). Hence, we will assume that $P_i$ and $Q_i$
are not of maximum rank. The simplest case here is given by matrices
of rank one which can be represented as
\[
P_i = u^{}_i {}^{t\!} w^{}_i, \qquad Q_i = x^{}_i {}^{t\!} y^{}_i,
\]
where $u$, $w$, $x$ and $y$ are $n$-dimensional column vectors. This
representation allows one to write the relations (\ref{e:3.27}) and
(\ref{e:3.28}) as
\begin{gather}
{}^{t\!} y_i + \sum_{j=1}^r \sum_{k=1}^n \frac{\nu_i}{\nu_i -
\epsilon_n^k \mu_j} ({}^{t\!}  y_i h^k u^{}_j) {}^{t\!} w_j h^{-k} =
0, \label{e:3.39} \\
u^{}_i + \sum_{j=1}^r \sum_{k=1}^n \frac{\mu_i}{\mu_i - \epsilon_n^k
\nu_j} h^k x^{}_j ({}^{t\!} y_j h^{-k} u^{}_i) = 0. \label{e:3.40}
\end{gather}
Using the identity
\begin{equation}
\sum_{k=0}^{n-1} \frac{z \epsilon_n^{-jk}}{z - \epsilon_n^k} = n
\frac{z^{n-|j|_n}}{z^n - 1},
\label{e:3.41}
\end{equation}
where $|j|_n$ is the residue of division of $j$ by $n$, we can rewrite
(\ref{e:3.39})
in components terms,
\begin{equation}
y_{i k} + n \sum_{j=1}^r (R_k)_{i j} w_{j k} = 0.
\label{e:3.42}
\end{equation}
Here the $r \times r$ matrices $R_k$ are defined as
\[
(R_k)_{i j} = \frac{1}{\nu_i^n - \mu_j^n} \sum_{\ell=1}^n \nu_i^{n -
|\ell-k|_n} \mu_j^{|\ell-k|_n} y_{i \ell} u_{j \ell}.
\]
The same identity (\ref{e:3.41}) allows one to write component form
of (\ref{e:3.40}) as
\begin{equation}
u_{i k} + n \sum_{j=1}^r x_{j k} (S_k)_{j i} = 0, \label{e:3.43}
\end{equation}
where
\[
(S_k)_{j i} = - \frac{1}{\nu^n_j - \mu^n_i} \sum_{\ell=1}^n
\nu_j^{|k-\ell|_n} \mu_i^{n - |k-\ell|_n} y_{j \ell} u_{i \ell}.
\]
With the help of the equality
\[
n - 1 - |i-1|_n = |-i|_n
\]
it is straightforward to demonstrate that
\[
(S_k)_{j i} = - \frac{\mu_i}{\nu_j} (R_{k+1})_{j i}.
\]
Consequently, we can write the equation (\ref{e:3.43}) as
\begin{equation}
u_{i k} - n \mu_i \sum_{j=1}^r x_{j k} \frac{1}{\nu_j} (R_{k+1})_{j i}
= 0.
\label{e:3.44}
\end{equation}
We use the equations (\ref{e:3.42}) and (\ref{e:3.44}) to express the
vectors $w_i$ and $x_i$ via the vectors $u_i$ and $y_i$,
\[
w_{i k} = - \frac{1}{n} \sum_{j=1}^r (R^{-1}_k)_{i j} y_{j k}, \qquad
x_{i k} = \frac{1}{n} \sum_{j=1}^r u_{j k} \frac{1}{\mu_j} (R^{-1}
_{k+1})_{j i} \nu_i.
\]
As the result, we come to the following solution of the relations
(\ref{e:3.27}) and (\ref{e:3.28}):
\[
(P_i)_{k \ell} = - \frac{1}{n} u_{i k} \sum_{j=1}^r (R^{-1}_\ell)_{i
j} y_{j \ell}, \qquad
(Q_i)_{k \ell} = \frac{1}{n} \sum_{j=1}^r u_{j k} \frac{1}{\mu_j}
(R^{-1}_{k+1})_{j i}
\nu_i y_{i \ell}.
\]

Further, it follows from (\ref{e:3.39}) and (\ref{e:3.40}) that, to
fulfill also (\ref{e:3.31})--(\ref{e:3.34}), it is sufficient to
satisfy the equations
\begin{gather}
\partial_- y_i = \nu_i^{-1} {}^{t\!} c_- y_i, \qquad \partial_+ y_i =
\nu_i^{} {}^{t\!} c_+ y_i, \label{e:3.45} \\
\partial_- u_i = - \mu_i^{-1} c_- u_i, \qquad \partial_+ u_i = -
\mu_i^{} c_+ u_i. \label{e:3.46}
\end{gather}
The $n$-dimensional column vectors $\theta_\rho$, defined by the
relation (\ref{e:3.13}), are eigenvectors of the matrices ${}^{t\!}
c_-$, ${}^{t\!} c_+$, $c_-$ and $c_+$,
\[
{}^{t\!} c_- \theta_\rho = m \epsilon_n^\rho \theta_\rho, \qquad {}
^{t\!} c_+ \theta_\rho = m \epsilon_n^{-\rho} \theta_\rho, \qquad c_-
\theta_\rho = m \epsilon_n^{-\rho} \theta_\rho, \qquad c_+ \theta_\rho
= m \epsilon_n^\rho \theta_\rho,
\]
and form a basis in the space $\bbC^n$. Hence, the general solution of
the equations (\ref{e:3.45}) and (\ref{e:3.46}) can be written in the
form
\[
u_{i k} = \sum_{\rho=1}^n c_{i \rho} \: \epsilon_n^{k \rho} \: \rme^{-
Z_\rho(\mu_i)}, \qquad
y_{i k} = \sum_{\rho=1}^n d_{i \rho} \: \epsilon_n^{k \rho} \:
\rme^{Z_{-\rho}(\nu_i)},
\]
where $c_{i \rho}$, $d_{i \rho}$ are arbitrary constants and
\[
Z_\rho(\mu) = m (\epsilon_n^{-\rho} \mu^{-1} z^- + \epsilon_n^\rho \mu
\: z^+).
\]

Thus, we see that it is possible to satisfy (\ref{e:3.27}),
(\ref{e:3.28}) and (\ref{e:3.31})--(\ref{e:3.34}). This gives us
solutions of the Toda equations (\ref{e:2.27}). Let us show that they
can be written in a simple determinant form.

Using (\ref{e:3.38}) and (\ref{e:3.25}), one gets
\[
\gamma = \psi_\infty = I_n + \sum_{i = 1}^r \sum_{k=1}^n h^k P_i h^{-
k}.
\]
For the matrix elements of $\gamma$ this gives the expression
\[
\gamma_{k \ell} = \delta_{k \ell} \left( 1 + n \sum_{i = 1}^r (P_i)
_{kk} \right)
= \delta_{k \ell} \left( 1 - \sum_{i, j = 1}^r u_{i k}(R^{-1}_k)_{i j}
y_{j k} \right).
\]
Hence, we have
\[
\Gamma_\alpha = 1 - \sum_{i, j = 1}^r u_{i \alpha} (R^{-1}_\alpha)_{i
j} y_{j \alpha}.
\]
To this expression can also be given the form
\[
\Gamma_\alpha = 1 - {}^{t\!} u^{}_\alpha R^{-1}_\alpha y^{}_\alpha,
\]
where $u_\alpha$ and $y_\alpha$ are $r$-dimensional column vectors
with the components $u_{i \alpha}$ and $y_{i \alpha}$, respectively.

We assume for convenience that the functions $u_{i \alpha}$ and $y_{i
\alpha}$ are defined for arbitrary integral values of $\alpha$ and
\[
u_{i, \alpha + n} = u_{i \alpha}, \qquad y_{i, \alpha + n} = y_{i
\alpha}.
\]
The periodicity of $R_\alpha$ in the index $\alpha$ follows from the
definition. It appears that it is more appropriate to use quasi-
periodic quantities $\wt u_\alpha$, $\wt y_\alpha$ and $\wt R_\alpha$
defined as
\begin{gather*}
\wt u_\alpha = M^\alpha u_\alpha, \qquad \wt y_\alpha = N^{-\alpha}
y_\alpha, \\
\wt R_\alpha = N^{-\alpha} R_\alpha M^\alpha,
\end{gather*}
where $N$ and $M$ are diagonal $r \times r$ matrices given by
\[
N_{i j} = \nu_i \delta_{i j}, \qquad M_{i j} = \mu_i \delta_{i j}.
\]
For these quantities one has quasi-periodicity conditions
\begin{gather*}
\wt u_{\alpha+n} = M^n \wt u_\alpha, \qquad \wt y_{\alpha+n} = N^{-n}
\wt y_\alpha, \\
\wt R_{\alpha+n} = N^{-n} \wt R_\alpha M^n.
\end{gather*}
The expression of the matrix elements of the matrices $\wt R_\alpha$
through the functions $\wt y_{i \alpha}$ and $\wt u_{i \alpha}$
has a remarkably simple form \cite{Mik81}
\begin{equation}
(\wt R_\alpha)_{i j} = \frac{1}{\nu_i^n - \mu_j^n} \left( \mu_j^n
\sum_{\beta=1}^{\alpha-1} \wt y_{i \beta} \wt u_{j \beta} + \nu_i^n
\sum_{\beta=\alpha}^n \wt y_{i \beta} \wt u_{j \beta} \right).
\label{e:3.47a}
\end{equation}
In terms of the quasi-periodic quantities, for the functions
$\Gamma_\alpha$
we have
\[
\Gamma_\alpha = 1 - {}^{t\!} \wt u^{}_\alpha \wt R^{-1}_\alpha \wt
y^{}_\alpha,
\]
and it can be shown that
\[
\Gamma_\alpha = \frac{\det(\wt R^{}_\alpha - \wt y^{}_\alpha \: {}
^{t\!} \wt u_\alpha)}
{\det \wt R^{}_\alpha}.
\]
Using the explicit form of $\wt R_\alpha$, one comes to the equality
\[
\wt R^{}_{\alpha+1} = \wt R^{}_\alpha - \wt y^{}_\alpha \: {}^{t\!}
\wt u_\alpha,
\]
Therefore, one can write \cite{Mik81}
\begin{equation}
\Gamma_\alpha = \frac{\det \wt R_{\alpha+1}}{\det \wt R_\alpha}.
\label{e:3.47}
\end{equation}

\subsection{Solitons through the rational dressing} \label{s:3.3}

To obtain a one-soliton solution one puts $r=1$. In this case $\wt
R_\alpha$ are ordinary functions for which one has the expression
\[
\wt R_\alpha = \frac{1}{\nu^n - \mu^n} \sum_{\rho, \sigma = 1}^n
c_\rho \: d_\sigma \: \rme^{-Z_\rho(\mu) + Z_{-\sigma}(\nu)}
\left[ \mu^n \sum_{\beta=1}^{\alpha-1} \mu^\beta \: \nu^{-\beta}
\epsilon_n^{(\rho + \sigma) \beta} + \nu^n \sum_{\beta = \alpha}^n
\mu^\beta \: \nu^{-\beta} \epsilon_n^{(\rho + \sigma) \beta} \right].
\]
It is not difficult to verify that
\[
\mu^n \sum_{\beta = 1}^{\alpha - 1} \mu^\beta \: \nu^{-\beta}
\epsilon_n^{(\rho + \sigma) \beta} + \nu^n \sum_{\beta = \alpha} ^n
\mu^\beta \: \nu^{-\beta} \epsilon_n^{(\rho + \sigma) \beta} = (\nu^n
- \mu^n) \: \mu^\alpha \: \nu^{-\alpha} \frac{\epsilon_n^{(\rho +
\sigma) \alpha}}{1 - \mu \: \nu^{-1} \epsilon_n^{\rho + \sigma}}.
\]
Thus one obtains the following expression for $\wt R_\alpha$:
\[
\wt R_\alpha = \mu^\alpha \: \nu^{-\alpha} \sum_{\rho, \sigma = 1}^n
c_\rho \: d_\sigma \: \rme^{-Z_\rho(\mu) + Z_{-\sigma}(\nu)}
\frac{\epsilon_n^{(\rho + \sigma) \alpha}}{1 - \mu \: \nu^{-1}
\epsilon_n^{\rho + \sigma}}.
\]

To obtain a solution which depends on only one combination of $z^-$
and $z^+$ we suppose that $c_\rho$ is different from zero for only one
value of $\rho$ which we denote by $I$, and that $d_\sigma$ is
different from zero for only two values of $\sigma$ which we denote by
$J$ and $K$. In this case we arrive at a simplified version of $\wt
R_\alpha$, that is
\[
\wt R_\alpha = \mu^\alpha \: \nu^{-\alpha} c_I \: \rme^{-Z_I(\mu)}
\left[ d_J \: \rme^{Z_{-J}(\nu)} \frac{\epsilon_n^{(I + J) \alpha}}{1
- \mu \: \nu^{-1} \epsilon_n^{I + J}} + d_K \: \rme^{Z_{-K}(\nu)}
\frac{\epsilon_n^{(I + K) \alpha}}{1 - \mu \: \nu^{-1} \epsilon_n^{I +
K}} \right],
\]
and the corresponding solution can be written as
\[
\Gamma_\alpha = \mu \: \nu^{-1} \epsilon_n^{I+J} \: \frac{1 + d \:
\epsilon_n^{(K-J)(\alpha + 1)}
\: \rme^{Z_{-K}(\nu) - Z_{-J}(\nu)}}{1 + d \: \epsilon_n^{(K - J)
\alpha} \: \rme^{Z_{-K}(\nu) - Z_{-J}(\nu)}},
\]
where
\[
d = \frac{d_K(1 - \mu \: \nu^{-1} \epsilon_n^{I+J})}{d_J(1 - \mu \:
\nu^{-1} \epsilon_n^{I+K})}.
\]
Making use of the freedom in multiplying a solution by a constant, we
can write the obtained solution as (\ref{e:3.18}),  where $\rho = K-
J$, $\kappa_\rho$ is defined by (\ref{e:3.16}), $\zeta = - \rmi
\epsilon_n^{-(K+J)/2} \nu$, and $\delta$ is a constant introduced by
the relation $\exp \delta = d$. Thus we arrive at the one-soliton
solution obtained before by the Hirota's method.

In the case of $r > 1$ (multi-soliton construction) we suppose that
for any $i$ the coefficients $c_{i \rho}$ are different from zero for
only one value of $\rho$ which we denote by $I_i$, and that the
coefficients $d_{i \sigma}$ are different from zero for only two
values of $\sigma$ which we denote by $J_i$ and $K_i$. This leads to
the following expression for the matrix elements of the matrices $\wt
R_\alpha$:
\begin{multline*}
(\wt R_\alpha)_{i j} = \nu_i^{-\alpha} \epsilon_n^{J_i \alpha} d_{J_i}
\rme^{Z_{-{J_i}}(\nu_i)}
\\ \times \left[ \frac{1}{1 - \mu_j^{} \: \nu_i^{-1} \epsilon_n^{I_j +
J_i}}
+ \frac{d_{K_i}}{d_{J_i}} \: \rme^{Z_{- K_i}(\nu_i) - Z_{-J_i}(\nu_i)}
\frac{\epsilon_n^{(K_i - J_i) \alpha}}{1 - \mu_j^{} \: \nu_i^{-1}
\epsilon_n^{I_j + K_i}} \right] \mu_j^\alpha \: \epsilon_n^{I_j
\alpha} c_{I_j} \: \rme^{- Z_{I_j}(\mu_j)}.
\end{multline*}
Immediately we see from (\ref{e:3.47}) that the solution in question
has
the form
\begin{equation}
\Gamma_\alpha = \left[ \prod_{i = 1}^r \mu_i^{} \nu_i^{-1}
\epsilon_n^{I_i + J_i} \right]
\frac{\det \wt R'_{\alpha + 1}}{\det \wt R'_\alpha},
\label{e:3.48}
\end{equation}
where the matrices $\wt R'_\alpha$ are defined by
\[
(\wt R'_\alpha)_{i j} = \frac{1} {1 - \mu_j^{} \: \nu_i^{-1}
\epsilon_n^{I_j + J_i}}
+ \frac{d_{K_i}}{d_{J_i}} \: \rme^{Z_{- K_i}(\nu_i) - Z_{-J_i}(\nu_i)}
\frac{\epsilon_n^{(K_i - J_i) \alpha}}{1 - \mu_j^{} \: \nu_i^{-1}
\epsilon_n^{I_j + K_i}}.
\]
Using the matrices $D$ defined in appendix \ref{apa}, we rewrite the
expression for $\wt R'_\alpha$ in the form
\[
(\wt R'_\alpha)_{i j} =  D_{i j} (\nu \epsilon_n^{-J}, \mu
\epsilon_n^I) + \frac{d_{K_i}}{d_{J_i}} \: \epsilon_n^{(K_i - J_i)
\alpha} \rme^{Z_{- K_i}(\nu_i) - Z_{-J_i}(\nu_i)} D_{i j}(\nu
\epsilon_n^{-K}, \mu \epsilon_n^I).
\]
It is clear that instead of $\wt R'_i$ one can use in the
relation (\ref{e:3.48}) the matrices $\wt R''_i$ defined as
\[
(\wt R''_\alpha)_{i j} = \delta_{i j}  + \frac{d_{K_i}}{d_{J_i}} \:
\epsilon_n^{(K_i - J_i) \alpha}  \rme^{Z_{-K_i}(\nu_i) - Z_{-J_i}
(\nu_i)} \sum_{k = 1}^r D_{i k}(\nu \epsilon_n^{-K}, \mu \epsilon_n^I)
D^{-1}_{k j} (\nu \epsilon_n^{-J}, \mu \epsilon_n^I).
\]
Using the equality (\ref{e:A.3}), one comes to the expression
\begin{equation}
(\wt R''_\alpha)_{i j} = \frac{\displaystyle \nu_i
\prod_{\substack{\ell = 1 \\ \ell \ne i}}^r (\nu_i \epsilon_n^{-J_i} -
\nu_\ell \epsilon_n^{-J_\ell})} {\displaystyle \epsilon_n^{J_i}
\prod_{\ell = 1}^r (\nu_i \epsilon_n^{-J_i} - \mu_\ell
\epsilon_n^{I_\ell})}
(T_\alpha)_{i j} \frac{\displaystyle \epsilon_n^{J_j} \prod_{\ell = 1}
^r (\nu_j \epsilon_n^{-J_j} - \mu_\ell \epsilon_n^{I_\ell})}
{\displaystyle \nu_j \prod_{\substack{\ell = 1 \\ \ell \ne j}}^r
(\nu_j \epsilon_n^{-J_j} - \nu_\ell \epsilon_n^{-J_\ell})},
\label{e:3.49}
\end{equation}
where
\[
(T_\alpha)_{i j} = \delta_{i j} + d_i \epsilon_n^{(K_i - J_i) \alpha}
\rme^{Z_{- K_i}(\nu_i) - Z_{-J_i}(\nu_i)} \frac{\nu_i \epsilon_n^{-
K_i} - \nu_i \epsilon_n^{-J_i}} {\nu_i \epsilon_n^{-K_i} - \nu_j
\epsilon_n^{-J_j}}
\]
and, with a slight abuse of notation,
\[
d_i = \frac{\displaystyle d_{K_i} \epsilon_n^{J_i}
\prod_{\substack{\ell = 1 \\ \ell \ne i}}^r (\nu_i \epsilon_n^{-K_i} -
\nu_\ell \epsilon_n^{-J_\ell}) \prod_{\ell = 1}^r
(\nu_i \epsilon_n^{-J_i} - \mu_\ell \epsilon_n^{I_\ell})}
{\displaystyle d_{J_i} \epsilon_n^{K_i}
\prod_{\substack{\ell = 1 \\ \ell \ne i}}^r (\nu_i \epsilon_n^{-J_i} -
\nu_\ell \epsilon_n^{-J_\ell}) \prod_{\ell = 1}^r (\nu_i \epsilon_n^{-
K_i} - \mu_\ell \epsilon_n^{I_\ell})}.
\]
Utilizing the expression (\ref{e:3.49}) and having in mind the freedom
in multiplying a solution by a constant, we write the solution under
consideration as follows:
\[
\Gamma_\alpha = \frac{\det T_{\alpha+1}}{\det T_\alpha}.
\]
Defining $\rho_i = K_i - J_i$, $\zeta_i = - \rmi \epsilon_n^{-(K_i +
J_i)/2} \nu_i$ and introducing constants $\delta_i$ satisfying the
relations $\exp \delta_i = d_i$, one can write
\begin{equation}
(T_\alpha)_{i j} = \delta_{i j} + \epsilon_n^{\rho_i \alpha} \exp[{m
\kappa_{\rho_i}
(\zeta_i^{-1} z^- + \zeta_i z^+) + \delta_i}] \,
\frac{\epsilon_n^{-\rho_i/2} \zeta_i - \epsilon_n^{\rho_i/2} \zeta_i}
{\epsilon_n^{-\rho_i/2} \zeta_i - \epsilon_n^{\rho_j/2}\zeta_j}.
\label{e:3.50}
\end{equation}
It is proved in appendix \ref{apb} that
\begin{equation}
\det T_{\alpha+1} = 1 + \sum_{i = 1}^r E_{\alpha i} + \sum_{\ell = 2}
^r \left[ \sum_{1 \le i_1 < i_2 < \ldots < i_\ell \le r}
\left(\prod_{1 \le j < k \le \ell} \eta_{i_j i_k} \right) E_{\alpha
i_1} E_{\alpha i_2} \ldots E_{\alpha i_\ell} \right], \label{e:3.51}
\end{equation}
where the functions $E_{\alpha i}$ and $\eta_{i_j i_k}$ are defined by
the relations (\ref{e:3.15}) and (\ref{e:3.19}) respectively. Thus, we
come to the multi-soliton solutions which coincide with those obtained
by the Hirota's method. The Hirota's $\tau$-functions (\ref{e:3.20})
are given by the equality
\[
\tau_\alpha = \det T_{\alpha + 1}.
\]
It is clear that the quantities $\eta_{i_j i_k}$ here make the same
sense as do the normal ordering coefficients effectively describing
the interaction between solitons in the vertex operators approach of
Olive, Turok and Underwood \cite{OliTurUnd92, OliTurUnd93}. We refer
the reader to the papers \cite{ZhuCal93, BeJo97, BeJo98} for some more
specific properties of such coefficients.

\section{Conclusion}

In this paper we have considered abelian Toda systems associated with
the loop groups of the complex general linear groups. We have reviewed
two different approaches to construct soliton solutions to these
equations in the untwisted case, namely, the Hirota's and rational
dressing methods. Subsequently, basic ingredients representing soliton
solutions within the frameworks of these methods have been explicitly
related. As we have seen in section \ref{s:3.2}, the rational dressing 
method allows one to construct solutions to the loop Toda equations, 
presenting them as the ratio of the determinants of specific matrices 
(\ref{e:3.47a}), (\ref{e:3.47}), and they actually represent a class 
of solutions being wider than that formed by the soliton solutions of 
the Hirota's method in section \ref{s:3.1}: By setting the initial-data 
coefficients arising in the rational dressing method to some specific 
values we have shown in section \ref{s:3.3} that the Hirota's soliton 
solutions are contained among the solutions constructed by the rational
dressing approach.    

Note also that the reduction to the systems based on the loop
groups of the complex special linear groups can easily be performed.

Our consideration can be generalized to Toda systems based on other
loop groups, such as twisted loop groups of the complex general linear
groups, twisted and untwisted loop groups of the complex orthogonal
and symplectic groups. However, one should take into account that the
change of field variables in the Hirota's method is more tricky there,
and besides, when applying the rational dressing to obtain soliton
solutions, one faces that the pole positions of the dressing
meromorphic mappings and their inverse ones are to be related,
just due to the group conditions. These circumstances make part of the
formulae more intricate than in the general linear case considered in
the present paper. We will address to this problem and present our
results in some future publications.

\vskip2mm
This work was supported in part by the Russian Foundation for Basic
Research under grant \#07--01--00234.

\appendix

\section*{Appendix}

\setcounter{section}{1}
\setcounter{equation}{0}

\subsection{\mathversion{bold}Some properties of the matrices $D$}
\label{apa}

In this appendix we investigate $r \times r$ matrices $D(f, g)$ with
matrix elements given by the equality
\[
D_{i j} (f, g) = \frac{1}{1 - f_i^{-1} g_j^{}} = \frac{f_i}{f_i -
g_j}.
\]
Let us show that for the matrix elements of the inverse matrix
$D^{-1}(f,g)$ one has the representation
\begin{equation}
D_{i j}^{-1} (f, g)
= \frac{\displaystyle \prod_{\substack{\ell = 1 \\ \ell \ne j}}^r
(f_\ell - g_i)
\prod_{\ell = 1}^r (f_j - g_\ell)} {f_j \displaystyle
\prod_{\substack{\ell = 1 \\ \ell \ne i}}^r
(g_\ell - g_i) \prod_{\substack{\ell = 1 \\ \ell \ne j}}^r (f_j -
f_\ell)}.
\label{e:A.1}
\end{equation}
To prove the above equality one has to demonstrate that
\begin{equation}
\sum_{k=1}^r \frac{f_ i \displaystyle
\prod_{\substack{\ell = 1 \\ \ell \ne j}}^r
(f_ \ell - g_k) \prod_{\ell = 1}^r (f_j - g_\ell)}
{f_j (f_i - g_k) \displaystyle \prod_{\substack{\ell = 1 \\ \ell \ne
k}}^r ( g_\ell - g_k)
\prod_{\substack{\ell = 1 \\ \ell \ne j}}^r
(f_j - f_\ell)} = \delta_{i j}.
\label{e:A.2}
\end{equation}
Consider the set of meromorphic functions of $z$ defined as
\[
F_{i j}(f, g, z) = \frac{\displaystyle \prod_{\substack{\ell = 1 \\
\ell \ne j}}^r (f_\ell -
z)}{(f_i - z)\displaystyle \prod_{\ell = 1}^r (g_\ell - z)}.
\]
The residue of $F_{i j}(f, g, z)$ at infinity is equal to zero,
therefore, the sum of the residues at the point $f_i$ and at the
points $g_\ell$, $\ell = 1, \ldots, r$, is also zero. Hence we have
the following equality
\[
\sum_{k=1}^r \frac{\displaystyle \prod_{\substack{\ell = 1 \\ \ell \ne
j}}^r (f_\ell -
g_k)}{(f_i - g_k)\displaystyle \prod_{\substack{\ell = 1 \\ \ell \ne
k}}^r (g_\ell - g_k)} = {} -
\frac{\displaystyle \prod_{\substack{\ell = 1 \\ \ell \ne j}}^r
(f_\ell - f_i)}{\displaystyle
\prod_{\ell = 1}^r (g_\ell - f_i)},
\]
and, therefore,
\[
\sum_{k=1}^r \frac{f_i \displaystyle \prod_{\substack{\ell = 1 \\ \ell
\ne j}}^r (f_ \ell - g_k)
\prod_{\ell = 1}^r (f_j - g_\ell)} {f_j (f_i - g_k)
\displaystyle \prod_{\substack{\ell = 1 \\ \ell \ne k}}^r ( g_\ell -
g_k)
\prod_{\substack{\ell = 1 \\ \ell \ne j}}^r
(f_j - f_\ell)} = \frac{f_i \displaystyle
\prod_{\substack{\ell = 1 \\ \ell \ne j}}^r
(f_i - f_\ell) \prod_{\ell = 1}^r
(f_j - g_\ell)}{f_j \displaystyle \prod_{\ell = 1}^r
(f_i - g_\ell) \prod_{\substack{\ell = 1 \\ \ell \ne j}}^r (f_j -
f_\ell)}.
\]
Now, taking into account the identity
\[
\prod_{\substack{\ell = 1 \\ \ell \ne j}}^r (f_i - f_\ell) /
\prod_{\substack{\ell = 1 \\ \ell \ne j}}^r
(f_j - f_\ell) = \delta_{i j},
\]
we see that the relation (\ref{e:A.2}) is true. Thus the equivalent
relation (\ref{e:A.1}) is also true.

In a similar way one can prove the validity of the equality
\begin{equation}
\sum_{k = 1}^r D_{i k} (\wt f, g)
D^{-1}_{k j} (f, g) = \frac{\wt f_ i \displaystyle
\prod_{\substack{\ell = 1 \\ \ell \ne j}}^r
(\wt f_i - f_\ell) \prod_{\ell = 1}^r (f_j - g_\ell)}
{f_j \displaystyle \prod_{\ell = 1}^r
(\wt f_i - g_\ell)
\prod_{\substack{\ell = 1 \\ \ell \ne j}}^r
(f_j - f_\ell)}.
\label{e:A.3}
\end{equation}

\subsection{Proof of relation (\ref{e:3.51})}
\label{apb}

Proceeding from the relation (\ref{e:3.50}), one obtains
\[
(T_{\alpha+1})_{ij} = \delta_{ij} + E_{\alpha i} \, \frac{\tilde f_i -
f_i}{\tilde f_i - f_j},
\]
where
\begin{equation}
\wt f_i = \epsilon_n^{-\rho_i/2} \zeta_i, \qquad f_i =
\epsilon_n^{\rho_i/2} \zeta_i. \label{e:A.4}
\end{equation}
and the functions $E_{\alpha i}$ are defined by the relation
(\ref{e:3.15}). Then it is not difficult to get convinced that
\begin{equation}
\det T_{\alpha+1} = 1 + \sum_{i = 1}^r E_{\alpha i} + \sum_{\ell = 2}
^r \left[
\sum_{1 \le i_1 < i_2 < \ldots < i_\ell \le r} \eta_{i_1 i_2 \ldots
i_\ell} E_{\alpha i_1}
E_{\alpha i_2} \ldots E_{\alpha i_\ell} \right],
\label{e:A.5}
\end{equation}
where
\[
\eta_{i_1 \ldots i_\ell} = \sum_{\pi \in \mathrm S_\ell} \mathrm{sgn}
(\pi) \prod_{j = 1}^\ell \frac{\wt f_{i_j} - f_{i_j}}{\wt f_{i_j} -
f_{i_{\pi(j)}}}.
\]
As is customary, we denote by $\mathrm S_\ell$ the symmetric group on
the set $\{1, 2, \ldots, \ell\}$ and by $\mathrm {sgn}(\pi)$ the
signature of the permutation $\pi$.

For $\ell=2$ one has
\[
\eta_{i_1 i_2} = 1 - \frac{(\wt f_{i_1} - f_{i_1})(\wt f_{i_2} -
f_{i_2})}{(\wt f_{i_1} - f_{i_2})(\wt f_{i_2} - f_{i_1})} =
\frac{(f_{i_1} - f_{i_2})(\wt f_{i_2} - \wt f_{i_1})}{(\wt f_{i_1} -
f_{i_2}) (\wt f_{i_2} - f_{i_1})}.
\]
Using the definition (\ref{e:A.4}) of $f_i$ and $\tilde f_i$, we see
that the quantities $\eta_{i_1 i_2}$ coincide with the coefficients
$\eta_{i_1 i_2}$ defined by the relation (\ref{e:3.19}).

Let us prove by induction that
\begin{equation}
\eta_{i_1 i_2 \ldots i_\ell}
= \prod_{1 \le j < k \le \ell}
\eta_{i_j i_k}. \label{e:A.6}
\end{equation}
Certainly, for $\ell=2$ the equality (\ref{e:A.6}) is valid. Suppose
that it is valid up to some fixed value of $\ell$ and show that it is
valid for its value incremented by one.

The group $\mathrm S_\ell$ can be identified with a subgroup of
$\mathrm S_{\ell+1}$ formed by the permutations $\pi \in \mathrm
S_{\ell+1}$ satisfying the condition $\pi(\ell+1) = \ell+1$. Denote by
$\pi_m$, $m = 1,\ldots, \ell$, the transposition exchanging $m$ and
$\ell+1$ and represent the group $\mathrm S_{\ell+1}$ as the union of
the right cosets $\mathrm S_\ell \pi_m$. This allows us to write
\begin{equation}
\eta_{i_1 \ldots i_\ell i_{\ell+1}} = \eta_{i_1 \ldots i_\ell} -
\sum_{m=1}^\ell \sum_{\pi
\in \mathrm S_\ell} \mathrm{sgn}(\pi) \prod_{j=1}^{\ell+1} \frac{\wt
f_{i_j} - f_{i_j}}{\wt f_{i_j} - f_{i_{\pi(\pi_m(j))}}}. \label{e:A.7}
\end{equation}
It is not difficult to realize that
\[
\sum_{\pi \in \mathrm S_\ell} \mathrm{sgn}(\pi) \prod_{j=1}^{\ell+1}
\frac{\wt f_{i_j}
- f_{i_j}}{\wt f_{i_j} - f_{i_{\pi(\pi_m(j))}}} = \frac{(\wt f_{i_m} -
f_{i_m})(\wt f_{i_{\ell+1}}
- f_{i_{\ell+1}})}{(\wt f_{i_m} - f_{i_{\ell+1}}) (\wt f_{i_{\ell+1}}
- f_{i_m})} \left.
\eta_{i_1 \ldots i_\ell} \right|_{\wt f_{i_m} = \wt f_{i_{\ell+1}}},
\]
and that
\[
\left. \eta_{i_1 \ldots i_\ell} \right|_{\wt f_{i_m} = \wt
f_{i_{\ell+1}}} = \prod_{\substack{j = 1 \\ j \ne m}}^\ell \frac{(\wt
f_{i_{\ell+1}} - \wt f_{i_j}) (\wt f_{i_m} - f_{i_j})}{(\wt f_{i_m} -
\wt f_{i_j})(\wt f_{i_{\ell+1}} - f_{i_j})} \: \eta_{i_1 \ldots
i_\ell}.
\]
Using these equalities in (\ref{e:A.7}), we obtain
\begin{multline}
\eta_{i_1 \ldots i_\ell i_{\ell+1}}
= \eta_{i_1 \ldots i_\ell} \\*
+ \eta_{i_1 \ldots i_\ell} \frac{\displaystyle (\wt f_{i_{\ell+1}}
- f_{i_{\ell + 1}}) \prod_{j=1}^\ell (\wt f_{i_{\ell+1}} - \wt
f_{i_j})}{\displaystyle \prod_{j=1}^\ell(\wt f_{i_{\ell+1}} -
f_{i_j})} \sum_{m=1}^\ell \frac{\displaystyle \prod_{j=1}^\ell (\wt
f_{i_m} - f_{i_j})} {\displaystyle (\wt f_{i_m}- f_{i_{\ell+1}})
\prod_{\substack{j = 1 \\ j \ne m}}^{\ell + 1} (\wt f_{i_m} - \wt
f_{i_j})}.
\label{e:A.8}
\end{multline}
Now consider a meromorphic function of $z$ defined as
\[
F(f, \wt f, z) =\frac{\displaystyle \prod_{j=1}^\ell (z - f_{i_j})}
{\displaystyle (z - f_{i_{\ell+1}}) \prod_{j = 1}^{\ell+1} (z - \wt
f_{i_j})}.
\]
The equality of the sum of the residues of $F(f, \wt f, z)$ to zero
gives the relation
\[
\sum_{m=1}^\ell \frac{\displaystyle \prod_{j=1}^\ell (\wt f_{i_m} -
f_{i_j})} {\displaystyle (\wt f_{i_m}- f_{i_{\ell+1}})
\prod_{\substack{j = 1 \\ j \ne m}}^{\ell + 1} (\wt f_{i_m} - \wt
f_{i_j})} = - \frac{\displaystyle \prod_{j=1}^\ell (f_{i_{\ell+1}} -
f_{i_j})}{\displaystyle \prod_{j = 1}^{\ell + 1} (f_{i_{\ell + 1}} -
\wt f_{i_j})} - \frac{\displaystyle \prod_{j=1}^\ell  (\wt f_{i_{l +
1}} - f_{i_j})}{\displaystyle (\wt f_{i_{\ell+1}}- f_{i_{\ell+1}})
\prod_{j = 1}^\ell (\wt f_{i_{\ell + 1}} - \wt f_{i_j})}.
\]
Using it in (\ref{e:A.8}), we come to the equality
\[
\eta_{i_1 \ldots i_\ell i_{\ell + 1}} = \eta_{i_1 \ldots i_\ell}
\prod_{j = 1}^\ell \frac{(f_{i_j} - f_{i_{\ell + 1}})(\wt f_{i_{\ell +
1}} - \wt f_{i_j})}{(\wt f_{i_j} - f_{i_{\ell + 1}})
(\wt f_{i_{\ell + 1}} - f_{i_j})} = \eta_{i_1 \ldots i_\ell} \prod_{j
= 1}^\ell \eta_{i_j i_{\ell + 1}}
\]
that gives (\ref{e:A.6}). It is clear that (\ref{e:A.5}) and
(\ref{e:A.6}) prove the validity of (\ref{e:3.51}).

\end{document}